\documentclass[fleqn,10pt]{wlscirep}
\usepackage[utf8]{inputenc}
\usepackage[T1]{fontenc}

\title{Interplay between altermagnetism and superconductivity in two dimensions: intertwined symmetries and singlet-triplet mixing}

\author[1]{Kinga Jasiewicz}
\author[2]{Pawe{\l} W{\'o}jcik}
\author[1]{Micha{\l} Nowak}
\author[1,*]{Micha{\l} Zegrodnik}
\affil[1]{AGH University of Krakow, Academic Centre for Materials and Nanotechnology, al. A. Mickiewicza 30, 30-059 Krakow, Poland}
\affil[2]{AGH University of Krakow, Faculty of Physics and Applied Computer Science, al. A. Mickiewicza 30, 30-059 Krakow, Poland}

\affil[*]{michal.zegrodnik@agh.edu.pl}

\begin{abstract}
We study the interplay between altermagnetism and unconventional superconductivity for the case of two-dimensional square- and triangular-lattice systems. Our approach is based on an effective single particle Hamiltonian which mimics the alternating spin splitting characteristic for the $d$-$wave$ and $i$-$wave$ altermagnetic state. By supplementing the model with intersite pairing term we characterize the principal features of the coexistent altermagnetic-superconducting state as well as the possibility of inducing the Fulde-Ferrell-Larkin-Ovchinnikov (FFLO) phase. Our calculations show that the subtle interplay between the symmetries of the superconducting and altermagnetic order parameters as well as the shape/size of the Fermi surface 
lead to various types of anisotropic behaviors of the resultant non-zero momentum pairing,  which has not been possible in the originally proposed FFLO state. Moreover, in the considered systems additional pairing symmetries appear leading to an exotic multi-component order parameter with singlet-triplet mixing. To interpret the obtained data we analyze the Cooper pair density in the momentum space and the corresponding Fermi wave vector mismatch resulting from the altermagnetic spin splitting. We discuss our result in the context of possible applications like, e.g., the superconducting diode.
\end{abstract}

\begin{document}

\flushbottom
\maketitle
% * <john.hammersley@gmail.com> 2015-02-09T12:07:31.197Z:
%
%  Click the title above to edit the author information and abstract
%
\thispagestyle{empty}


\section*{Introduction}

Altermagnets (AM) have recently emerged as a promising new class of functional materials, drawing significant attention.~\cite{Smajkal_PRX_1,Smejkal_PRX_2,Mazin_PRX_Editorial,Mazin_Physics,Song2025,Yugui_2024_Review}. These systems are characterized by non-relativistic alternating spin polarization in momentum space together with no net spin imbalance. Altermagnetism is identified/proposed in various types of materials, including insulators~\cite{Ma2021_AM_insulator, Smajkal_PRX_1,Smejkal_PRX_2}, metals ~\cite{Elmers_Science_Metallic,Jiang2025_Metallic,Sarkar2025_Metallic}, semiconductors~\cite{Kim_PRL_Semiconducting,Zhu2024_Nature_Semiconducting}, 2D systems~\cite{Zhang2025_2D_RbV2_TeO2,jiang2024discoverymetallicroomtemperaturedwave_2D_dwave,Yu-Jun_2D_altermagnets} as well as superconductors~\cite{Smajkal_PRX_1,Smejkal_PRX_2,mazin2022notesaltermagnetismsuperconductivity}. Interestingly, different symmetries of alternating spin splitting may appear in momentum space defining the order of magnetization density. The $d$-$wave$, $g$-$wave$, and $i$-$wave$ symmetries have been considered so far ~\cite{Smajkal_PRX_1,Smejkal_PRX_2,Song2025,YuJun_Bilayer_AM}.

At the same time, the interplay between magnetism and superconductivity (SC) leads to a range of intriguing phenomena, both in terms of fundamental physics and applied perspectives~\cite{Eschrig_2015_SC_spintronics,Linder2015_SC_spintronics,Tero_SC_spintronics,Flensberg2021,Narita2022,Coronado2010,Hiroshi,Qimao}. The recent discovery of altermagnetic materials motivates further study in this direction. In particular, it has been proposed that alternating spin-splitting can induce the so-called Fulde-Ferrell-Larkin-Ovchinnikov state (FFLO) characterized by non-zero center-of-mass momentum of the Cooper pairs\cite{Zhang2024_SC_FFLO_NatCom,AnnicaBlack,Hermann2024,sumita2025,Sim_2025,Yusuke_2025}. A novelty of such a proposal with respect to previous considerations would be a possible formation of the FFLO state in the absence of any external magnetic field and/or net magnetization. This effect is also attractive in the context of the possible realization of an altermagnetic superconducting diode effect\cite{Mathias,AnnicaBlack_SC_diode,Fukaya_2025}. Some of the theoretical considerations focus on the interplay between a conventional $s$-$wave$ superconducting state with altermagnetism\cite{Zhang2024_SC_FFLO_NatCom,Sim_2025}. However, since the well-known parent cuprate material La$_2$CuO$_4$ has been proposed as a good candidate for realizing the altermagnetic phase, the unconventional $d$-$wave$ superconductivity, characteristic for the cuprates, has also gathered some attention\cite{Arun_AM_SC_t_J_model,AnnicaBlack,León2025}. In connection with this, organic compounds have also been analyzed together with some exotic form of the superconducting order parameter\cite{Hitoshi2023_organic_SC}. Since both the altermagnetic and superconducting states are characterized by various symmetries, an important general aspect of any analysis related to the SC and AM co-existence is the interplay between the symmetry factors themselves. 

Here, we carry out a comprehensive analysis focused on superconductivity and altermagnetism in two-dimensional systems by considering two selected lattices of significant importance: the square and triangular lattice. Our approach is based on an effective single particle Hamiltonian, which mimics the alternating spin splitting characteristic for the altermagnetic state. For the case of triangular lattice we propose a proper pattern of spin- and direction- dependent hoppings which reproduce the $i$-$wave$ altermagnetic symmetry of the resultant dispersion relations as well as the spin split Fermi surfaces. For the case of square lattice we employ the approach proposed in previous theoretical considerations\cite{AnnicaBlack} which is appropriate for the $d$-$wave$ altemagnetic symmetry. Additionally, we supplement our approach with an intersite pairing term, which allows us to model an unconventional superconducting state with various symmetries of the $\mathbf{k}$-dependent SC gap, including the $d$-$wave$ symmetry well known from the high-temperature superconductors as well as the $extended$ $s$-$wave$ symmetry discussed in the context of LAO/STO interfaces as well as highly overdoped cuprates\cite{Ziqiang_cuprate_s_w,ZHONG20161239,Zegrodni_2021_cuprate_s_w,Zegrodnik_2020_LAO_STO}. Nevertheless, in our study we do not limit ourselves to the two mentioned cases only. Instead, we take into account all possible pairing symmetries and their mixtures that are allowed in the selected square and triangular structures with nearest-neighbor pairing. Additionally, we consider the possibility of non-zero center-of-mass momentum of the Cooper pairs in all the considered cases and analyze the features of the resultant FFLO state. Our calculations show that the subtle interplay between the particular symmetries of the superconducting and altermagnetic order parameters as well as the shape/size of the Fermi surface determine the structure of the free energy minima appearing in the Cooper pair momentum space. Such an effect leads to various types of anisotropic behavior of the resultant non-zero momentum pairing, which has not been possible in the originally proposed FFLO\cite{Ferrell_FF,Larkin_LO}. Moreover, in the obtained FFLO state additional pairing symmetries appear, leading to an exotic form of the order parameter with singlet-triplet mixing. To interpret the obtained results we analyze the Cooper pair density in the momentum space and the corresponding Fermi wave vector mismatch resulting from the altermagnetic spin splitting. For the sake of completeness, we also show the effect of external magnetic field which introduces some non-zero net magnetization on top of alternating spin-splitting originating from the altermagnet.

%In particular, we focus on the $d$-$wave$, $extended$ $s$-$wave$, $d$+$id$-wave, $p$-$wave$) in the presence of the two mentioned altermagnetic orderings.

\section*{Model and method}
\begin{figure*}[ht]
    \centering   
        \includegraphics[scale=0.2]{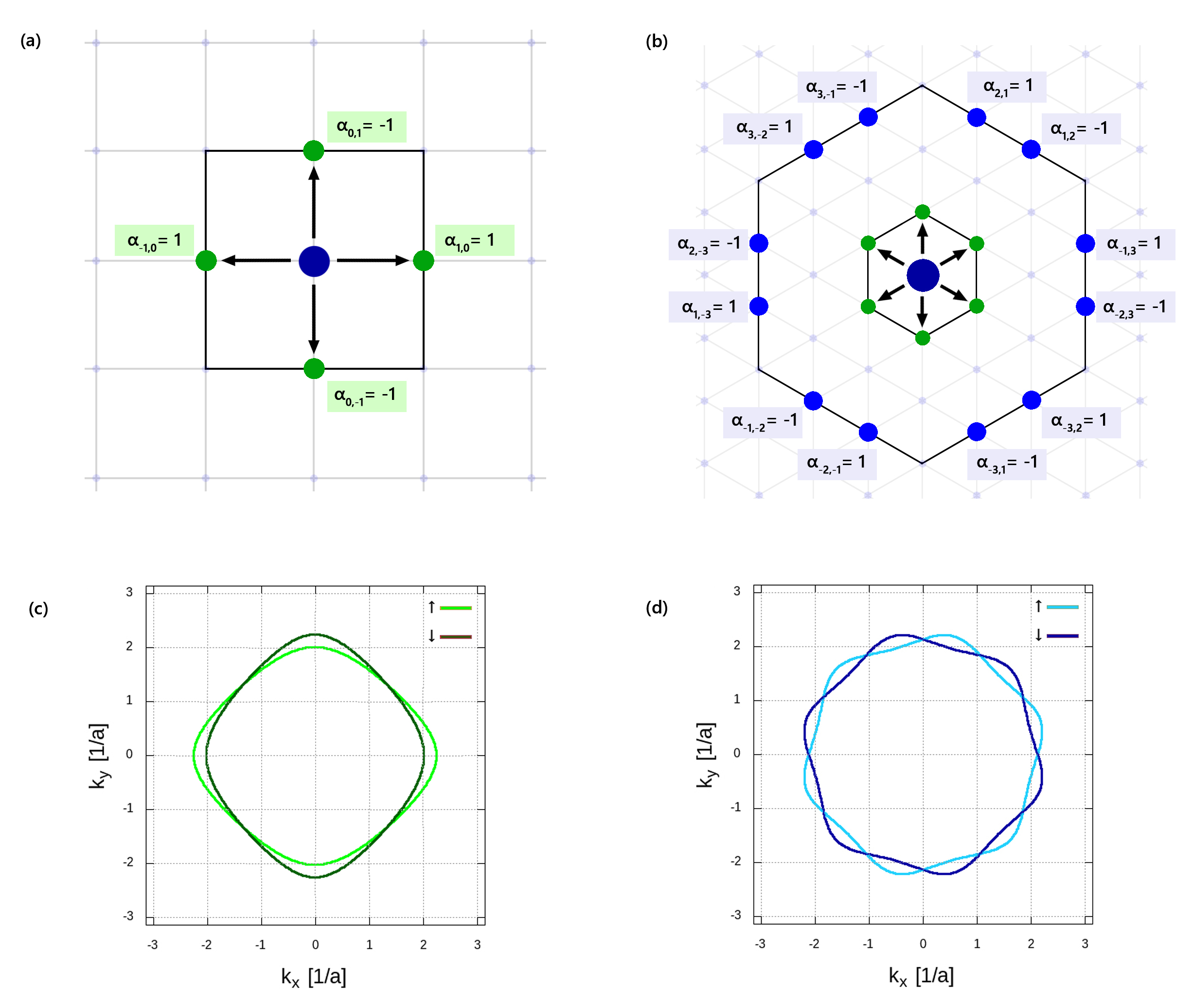}
    \caption{The values of the direction dependent $\alpha_{ij}$ factor appearing in Eq. (\ref{eq:H_t}) for nearest neighbor hopping on a square lattice which leads to the $d$-$wave$ altermagnetic symmetry (a) and for the third nearest neighbor hopping on a triangular lattice which leads to the $i$-$wave$ altermagnetic symmetry (b). Additionally in (a) and (b) we show the nearest neighbor hoppings marked by black arrows. In (c) and (d) we show the resultant spin split Fermi surfaces corresponding to the situations visualized in (a) and (b), respectively. 
    }
    \label{fig:theory_1}
\end{figure*}

We start with the effective model of the following form
\begin{equation}
    \hat{H}=\hat{H}_t+\hat{H}_{SC}\;,
    \label{eq:H_start}
\end{equation}
where $\hat{H}_t$ corresponds to the single particle part of the Hamiltonian which mimics the alternating spin splitting characteristic for the altermagnet, while $\hat{H}_{SC}$ is the intersite pairing term which introduces superconductivity of an unconventional type. The first term has the form
\begin{equation}
  \hat{H}_t = \sum_{ij \sigma} (t_{ij} + s_z\;\alpha_{ij}\;t_{\mathrm{am}}) \hat{a}^{\dagger}_{i\sigma} \hat{a}_{j\sigma}\;+\frac{\mu_BgB_z}{2}\sum_{i\sigma}s_z\;\hat{n}_{i\sigma},
  \label{eq:H_t}
\end{equation}
where $\hat{a}_{i\sigma}^{\dagger}$ ($\hat{a}_{i\sigma}$) creates (anihilates) electron with spin $\sigma$ at the $i$-th lattice site, $t_{ij}$ is the standard spin-independent part of the hopping between two lattice sites $j$ and $i$, while $s_z=1$ ($s_z=-1$) for $\sigma=\uparrow$ ($\sigma=\downarrow$), and $\alpha_{ij}=\pm 1$ is the direction dependent factor determining the symmetry of the altermagnetic state, to be defined explicitly below. The parameter $t_{\mathrm{am}}$ tunes the amplitude of the spin- and direction-dependent part of the hopping, defining the strength of the altermagnetic state. In this study we consider two types altermagnetic systems. Namely, a square lattice with $d$-$wave$ altermagnetic symmetry and a triangular lattice with $i$-$wave$ altermagnetic symmetry which are defined here in the following manner:
\begin{itemize}
\item \underline{Square lattice with $d$-$wave$ altermagnetic symmetry}: here we take both $t_{ij}$ and $\alpha_{ij}$ as non-zero to the nearest neighbors only, while $t_{ij}\equiv t$ and $\alpha_{ij}=\pm 1$ depending on the direction of the hopping as visualized in Fig. \ref{fig:theory_1}(a). The resultant Fermi surfaces with alternating $d$-$wave$ spin splitting are provided in Fig. \ref{fig:theory_1}(c),
\item \underline{Triangular lattice with $i$-$wave$ altermagnetic symmetry}: here we take into account the $t_{ij}$ as non-zero to the nearest neighbors only with $t_{ij}\equiv t$, however. in order to reconstruct the $i$-$wave$ symmetry we need to introduce non-zero $\alpha_{ij}=\pm 1$ parameter corresponding to the third nearest neighbors as shown in Fig. \ref{fig:theory_1}(b). The resultant Fermi surfaces with alternating $i$-$wave$ spin splitting are provided in Fig. \ref{fig:theory_1}(d). 
\end{itemize}
In both cases, the larger the $t_{\mathrm{am}}$ parameter is, the more significant the spin splitting is going to be; however, the directions in momentum space for which the spin splitting disappears are not going to be affected by the value of $t_{\mathrm{am}}$ itself since they originate from the particular symmetry of the aletrmagnetic state. For $t_{\mathrm{am}}=0$ no spin-spliting appears and the altermagnetic behavior is lost. For the sake of simplicity in all the cases considered we set $t=-1$ and analyze the physical properties of our solution as a function of $t_{\mathrm{am}}$ for the two cases described above. As an addition, we also investigate the interplay between external magnetic field and altermagnetism, therefore, we have supplemented our model with a Zeeman term corresponding to magnetic field oriented along the $z$-direction.

The explicit form of the pairing term from our starting Hamiltonian [Eq. (\ref{eq:H_start})] is the following
\begin{equation}
\hat{H}_{sc} =  - \frac{1}{2} J \sum_{ij\sigma} \hat{a}^{\dagger}_{i\sigma} \hat{a}^{\dagger}_{j\bar{\sigma}} \hat{a}_{j \bar{\sigma}} \hat{a}_{i\sigma}\;,
\label{eq:H_SC}
\end{equation}
where only nearest neighboring lattice sites are taken into account in the summation and $J$ determines the pairing strength. The $1/2$ factor is due to the fact that in the summation each bond between the lattice sites is taken into account twice. Therefore, to avoid double counting, we divide by two the entire term. Such pairing term corresponds to a generic form of an attractive interaction between the nearest neighbors that can lead to the creation of Cooper pairs with various symmetries of the $\mathbf{k}$-dependent SC gap. As we show in the next Section, in the absence of altermagnetic spin splitting such a term leads to the stability of even parity spin-singlet Cooper pairing. Nevertheless, the odd symmetries of the SC gap corresponding to the $S=0$ spin-triplet states are also considered here and appear to be nonzero in the presence of altermagnetic behavior. It should be noted that in both situations shown in Fig. \ref{fig:theory_1} for certain areas of the Fermi surface, there are no states with opposite momenta and spins available for pairing. Such Fermi wave vector mismatch has a detrimental effect on the stability of the superconducting state. However, it can be compensated by non-zero center of mass momentum of the Cooper pairs as we analyze in detail further on.

By applying the mean-field approximation to the pairing term, we obtain:			
\begin{equation}
	\hat{H}_{sc} = \frac{1}{2}  \sum_{ij\sigma} \big[ (\Delta_{ji}^{\bar{\sigma}\sigma})^{\ast} \hat{a}_{j\bar{\sigma}} \hat{a}_{i\sigma} + \Delta_{ji}^{\bar{\sigma}\sigma} \hat{a}_{i\sigma}^{\dagger} \hat{a}_{j\bar{\sigma}}^{\dagger} \big] + \frac{1}{2J} \sum_{ij \sigma} |\Delta_{ji}^{\bar{\sigma}\sigma}|^2,
    \label{eq:H_SC_HF_real_space}
\end{equation}
where we have introduced the real space pairing amplitudes 
\begin{equation}
  \Delta_{ji}^{\bar{\sigma}\sigma}=-J\langle\hat{c}_{j\bar{\sigma}}\hat{c}_{i\sigma}\rangle,\quad (\Delta_{ji}^{\bar{\sigma}\sigma})^{\ast}=-J\langle\hat{c}^{\dagger}_{i\sigma}\hat{c}^{\dagger}_{j\bar{\sigma}}\rangle.
  \label{eq:Delta_ji}
\end{equation}
In order to derive the self-consistent equations for the SC gaps we first transform the pairing part into momentum space
\begin{equation}
	\hat{H}_{sc} = \frac{1}{2} \sum_{\mathbf{k}\sigma}\Big( \Delta_{\mathbf{k}\mathbf{Q}}^{\bar{\sigma}\sigma}\;\hat{a}_{\mathbf{k}\sigma}^{\dagger}\hat{a}_{(-\mathbf{k}+\mathbf{Q})\bar{\sigma}}^{\dagger}  + (\Delta_{\mathbf{k}\mathbf{Q}}^{\bar{\sigma}\sigma})^{\ast}\; \hat{a}_{(-\mathbf{k}+\mathbf{Q})\bar{\sigma}}  \hat{a}_{\mathbf{k}{\sigma}} \Big) + \frac{1}{2J} \sum_{ij \sigma} |\Delta_{ji\mathbf{Q}}^{\bar{\sigma}\sigma}|^2,
    \label{eq:H_SC_HF_k_space}
\end{equation} 
%%%%%%%%%%%%%%%%%%%%%%%%%%%%%%%%%%%%%%%%%%%%%%%%%%%%%%%%%%%%%%%%%%%%%%%%%%%%%%
where we allow for the pairing between $(\mathbf{k},\;\sigma)$ and $(-\mathbf{k}+\mathbf{Q},\; \bar{\sigma})$ electrons, meaning that the total momentum of the Cooper pairs is $\mathbf{Q}$ which, in principle, can be non-zero possibly leading to the so-called Fulde-Ferrell (FF) phase. In such case the SC gap in momentum space has the form
\begin{equation}
    \Delta_{\mathbf{k}\mathbf{Q}}^{\bar{\sigma}\sigma} = \sideset{}{'}\sum_{i(j)} e^{i\mathbf{k}(\mathbf{R}_i-\mathbf{R}_j)} \Delta_{ji\mathbf{Q}}^{\bar{\sigma}\sigma}\;, 
    \label{eq:Delta_k_Q}
\end{equation}
where $\mathbf{R}_i$ and $\mathbf{R}_j$ vectors determine the positions of the $i$-th and $j$-th lattice sites and the summation runs over the $i$ index, which corresponds to the nearest neighbors of the $j$-th lattice site. Since we do not take into account any charge ordering the above summation does not depend on $j$. The $\mathbf{Q}$-dependent real space hopping amplitudes appearing in Eq. (\ref{eq:Delta_k_Q}) are defined as follows 
\begin{equation}
    \Delta_{ji\mathbf{Q}}^{\bar{\sigma}\sigma} = -\frac{J}{N} \sum_{\mathbf{k}} e^{-i\mathbf{k}(\mathbf{R}_i-\mathbf{R}_j)}\langle \hat{a}_{-\mathbf{k}+\mathbf{Q}\bar{\sigma}} \hat{a}_{\mathbf{k}\sigma} \rangle\;,
    \label{eq:Delta_ji_Q}
\end{equation}
where $N$ is the number of lattice sites in our system. For the sake of completeness, we also provide the relation between the $\mathbf{Q}$-dependent real space gap amplitude $\Delta_{ji\mathbf{Q}}^{\bar{\sigma}\sigma}$ defined by Eq. (\ref{eq:Delta_ji_Q}) and $\Delta_{ji}^{\bar{\sigma}\sigma}$ introduced by Eq. (\ref{eq:Delta_ji})
\begin{equation}
    \Delta_{ji\mathbf{Q}}^{\bar{\sigma}\sigma}=e^{i\mathbf{Q}\mathbf{R}_j}\Delta_{ji}^{\bar{\sigma}\sigma}.
\end{equation}
Within our approach, the Cooper pair momentum $\mathbf{Q}$ is determined by minimizing the energy of the system. More details related with the derivation of the pairing part in $\mathbf{k}$-space defined by Eq. (\ref{eq:H_SC_HF_k_space}) is provided in Appendix A.

In order to be able to write down the full Hamiltonian of the system in a compact matrix form, we derive the $\mathbf{k}$-space representation of the single-particle part. Namely,
\begin{equation}
\begin{split}
 \hat{H}_t &=  \sum_{\mathbf{k}\sigma} \epsilon_{\mathbf{k}\sigma} \hat{a}_{\mathbf{k}\sigma}^{\dagger} \hat{a}_{\mathbf{k}\sigma}=\frac{1}{2}\sum_{\mathbf{k}\sigma} \epsilon_{\mathbf{k}\sigma} \hat{a}_{\mathbf{k}\sigma}^{\dagger} \hat{a}_{\mathbf{k}\sigma} +\frac{1}{2}\sum_{\mathbf{k}\sigma} \epsilon_{(-\mathbf{k+Q})\sigma} \hat{a}_{(-\mathbf{k+Q})\sigma}^{\dagger} \hat{a}_{(-\mathbf{k+Q})\sigma}  \\
 &=\frac{1}{2}\sum_{\mathbf{k}\sigma} \epsilon_{\mathbf{k}\sigma} \hat{a}_{\mathbf{k}\sigma}^{\dagger} \hat{a}_{\mathbf{k}\sigma} -\frac{1}{2}\sum_{\mathbf{k}\sigma} \epsilon_{(-\mathbf{k+Q})\sigma} \hat{a}_{(-\mathbf{k+Q})\sigma} \hat{a}_{(-\mathbf{k+Q})\sigma}^{\dagger}+\frac{1}{2}\sum_{\mathbf{k}\sigma} \epsilon_{(-\mathbf{k+Q})\sigma}, 
 \label{eq:H_t_k_space}
 \end{split}
\end{equation}
where we have used the anticommutation relations and the fact that the system is translationally invariant. In the above equation, the explicit form of the dispersion relations is as follows
\begin{equation}
    \epsilon_{\mathbf{k}\sigma}=\sideset{}{''}\sum_{j(i)}(t_{ij} + s_z\;\alpha_{ij}\;t_{\mathrm{am}})e^{i\mathbf{k}(\mathbf{R}_i-\mathbf{R}_j)}+s_z\frac{\mu_0gB_z}{2}\;,
\end{equation}
where the summation runs over all the neighboring lattices sites $j$ surrounding lattice site $i$, for which the hopping amplitudes are taken to be non-zero as defined in the main text below Eq. (\ref{eq:H_t}).

Finally, by using Eqs. (\ref{eq:H_SC_HF_k_space}) and (\ref{eq:H_t_k_space}) we write down the full Hamiltonian of the system in $\mathbf{k}$ space with the possibility of non-zero momentum pairing
\begin{equation}
    \hat{H} = \frac{1}{2}\sum_{\mathbf{k}\sigma}\mathbf{\hat{h}}_{\mathbf{k}\sigma}^{\dagger}\mathbf{\hat{H}_{k\sigma}} \mathbf{\hat{h}}_{\mathbf{k\sigma}} + \frac{1}{2}\sum_{\mathbf{k}\sigma}\epsilon_{(-\mathbf{k}+\mathbf{Q})\sigma} + \frac{N}{2J} \sum_{i(j)\sigma} |\Delta_{ji\mathbf{Q}}^{\bar{\sigma}\sigma}|^2,
\end{equation}
where we have introduced the composite (2-dimensional) operators defined as
\begin{equation}
   \mathbf{\hat{h}}_{\mathbf{k}\sigma}^{\dagger} = 
    \begin{bmatrix}
        \hat{a}_{\mathbf{k}\sigma}^{\dagger} & \hat{a}_{(-\mathbf{k}+\mathbf{Q})\bar{\sigma}}\\
    \end{bmatrix},
    \label{eq:h_vector_composite}
\end{equation}
and $\mathbf{\hat{h}}_{\mathbf{k}\sigma}=(\mathbf{\hat{h}}_{\mathbf{k}\sigma}^{\dagger})^{\dagger}$, while the Hamiltonian matrix has the form
\begin{equation}
   \mathbf{\hat{H}_{k\sigma}} = 
    \begin{bmatrix}
        \epsilon_{\mathbf{k}\sigma}-\mu&    \Delta_{\mathbf{k}\mathbf{Q}}^{\bar{\sigma}\sigma} \\
        (\Delta_{\mathbf{k}\mathbf{Q}}^{\bar{\sigma}\sigma})^*  & -\epsilon_{(-\mathbf{k}+\mathbf{Q})\bar{\sigma}} -\mu \\
    \end{bmatrix}\;.
    \label{eq:H_matrix}
\end{equation}
By diagonalizing the above matrix and using Eq. (\ref{eq:Delta_ji_Q}) one can derive the self-consistent equations for the SC gap amplitudes between all four (six) nearest neighbors of the square (triangular) lattice in a standard manner by applying the Bogoliubov-de Gennes formalism. Note that for the case of the so-called Larking-Ovchinnikov state (LO) where two opposite momenta of the Cooper pairing are considered ($\mathbf{Q}$ and -$\mathbf{Q}$), a much more complicated situation takes place as the matrix Hamiltonian cannot be expressed in the block diagonal form shown above.

%Such an approach is analogous to the Bogolubov-de Gennes formalism introduced for the BCS theory. However, here an unconventional real-space pairing scenario is considered with the possibility of non-zero Cooper pairing in the presence of altermagnetic ordering. 

In our considerations, we do not limit ourselves to one specific symmetry of the SC gap. Instead, we consider all possible pairing symmetries that are allowed for the particular lattice at hand assuming only the pairing mechanism of the form given by Eq. (\ref{eq:H_SC}). This requires solving the set of self-consistent equations for all nearest-neighbor gap amplitudes without any additional constraints. Moreover, we do not assume that $\Delta_{ij\mathbf{Q}}^{\uparrow\downarrow}=-\Delta_{ij\mathbf{Q}}^{\downarrow\uparrow}$. We allow for $|\Delta_{ij\mathbf{Q}}^{\uparrow\downarrow}|\neq|\Delta_{ij\mathbf{Q}}^{\downarrow\uparrow}|$ which means that potential singlet-triplet mixing is also taken into account.

After all the nearest-neighbor gap parameters are calculated with the use of the self-consistent equations we determine the symmetry resolved SC gaps by using the following equation\cite{Zegrodnik_WSe2_1,Zegrodnik_WSe2_2}
\begin{equation}
    \Delta^{\sigma\bar{\sigma}}_{M,p}=\frac{i^p}{Z}\sum_{j(i)} e^{-iM\theta_{ji}}\Delta^{\sigma\bar{\sigma}}_{ji\mathbf{Q}},
    \label{eq:gap_symmetry_resolved}
\end{equation}
where the summation runs over the nearest neighbor lattice sites $j$ surrounding the site $i$, $Z$ is the coordination number of a given structure, and $\theta_{ji}=l\;2\pi/Z$ for $l=0,1,...,Z-1$ corresponds to subsequent angles between the positive half-$x$ axis and the $\mathbf{R}_{ji}=\mathbf{R}_j-\mathbf{R}_i$ vector, respectively. $M$ is the symmetry factor, which takes integer values and corresponds to possible pairing symmetries (cf. Table \ref{tab:symmetries_SC}). The $p$ parameter corresponds to the parity of the particular symmetry: for even parity symmetries we have $p=0$, and for odd parities we have $p=1$.

\begin{table}[!h]
  \centering
  \begin{minipage}[t]{0.45\textwidth}
    \centering
    \textbf{Triangular lattice ($Z=6$)} \\[0.5ex]
   \begin{tabular}{ c|c|c|c|c } 
 \hline\hline
 M & p & gap symmetry & parity & spin state \\ 
 \hline
 0 & 0 & extended $s$ & even & singlet \\ 
 \hline
 1 & 1  & $p_x+i\;p_y$ & odd & triplet \\ 
 \hline
 2 & 0  & $d_{x^2-y^2}+i\;d_{xy}$ & even & singlet \\ 
 \hline
 3 & 1  & $f$ & odd & triplet \\ 
 \hline
 4 & 0  & $d_{x^2-y^2}-i\;d_{xy}$ & even & singlet \\ 
 \hline
 5 & 1  & $p_x-i\;p_y$ & odd & triplet \\ 
 \hline\hline
 \end{tabular}
  \end{minipage}
  \hfill
  \begin{minipage}[t]{0.45\textwidth}
    \centering
    \textbf{Square lattice ($Z=4$)} \\[0.5ex]
\begin{tabular}{ c|c|c|c|c } 
 \hline\hline
 M & p & gap symmetry & parity & spin state \\ 
 \hline
 0 & 0 & extended $s$ & even & singlet \\ 
 \hline
 1 & 1  & $p_x+i\;p_y$ & odd & triplet \\ 
 \hline
 2 & 0  & $d_{x^2-y^2}$ & even & singlet \\ 
 \hline
 3 & 1  & $p_x-i\;p_y$ & odd & triplet \\ 
 \hline\hline
 \end{tabular}
  \end{minipage}
    \caption{The values of the symmetry factor M (first column) and the corresponding parity factor p (second column), appearing in Eq. (\ref{eq:gap_symmetry_resolved}) defining the possible pairing symmetries for the case of triangular lattice (left) and square lattice (right) with real-space nearest-neighbor pairing mechanism.}
    \label{tab:symmetries_SC}
\end{table}

Finally, one can write down the expressions for the real-space singlet and triplet gap amplitudes in the correlated state, which play the central role in the subsequent analysis
\begin{equation}
\begin{split}
    \Delta^{s}_{M,p}&=(\Delta^{\uparrow\downarrow}_{M,p}-\Delta^{\downarrow\uparrow}_{M,p})/\sqrt{2},\\
    \Delta^{t}_{M,p}&=(\Delta^{\uparrow\downarrow}_{M,p}+\Delta^{\downarrow\uparrow}_{M,p})/\sqrt{2}.\\
    \label{eq:gap_symmetries_spin}
\end{split}
\end{equation}
For the sake of clarity, in the following Sections, we use the symmetry names ($p\pm ip$, $d\pm id$, $f$ etc.) in the subscripts of the symmetry resolved superconducting gaps instead of the values of the $M$ and $p$ factors.

Another quantity which will be useful during our analysis is the pair density in $\mathbf{k}$-space which in our case of non-zero momentum pairing takes the form
\begin{equation}
    \gamma^{\bar{\sigma}\sigma}_{\mathbf{kQ}}=|\langle\hat{a}_{(\mathbf{-k+Q})\bar{\sigma}}\hat{a}_{\mathbf{k}\sigma}\rangle|.
    \label{eq:pair_density}
\end{equation}

Within our numerical scheme, we supplement the set of self-consistent equations for the SC gap parameters by an equation for the chemical potential as we treat the band filling as a parameter of our calculations. The band filling in this study is defined as the expectation value of the number of electrons per lattice site $n=\sum_{\sigma}\langle\hat{n}_{i\sigma} \rangle$, therefore, $n=0$, $n=1$, and $n=2$ correspond to empty, half-filled, and fully-filled band, respectively.

%It should be noted that while carrying out the transformation to reciprocal space we have assumed that the superconducting condensate can realize only one Cooper pair total momentum, which corresponds to the so-called Fulde-Ferrell (FF) phase. 
%This is in contrast to the concept introduced by Larkin and Ovchinnikov for which two $\mathbf{Q}$-vectors with opposite directions have been considered. 

%%%%%%%%%%%%%%%%%%%%%%%%%%%%%%%%%%%%%%%%%%%%%%%%%%%%%%%%%%%%%%%%%%%%%%%%%%%%%%%%%%%%%%%%%%%%

\section*{Results}
We have divided our analysis into two main parts. The first one corresponds to the square lattice with $d$-$wave$ altermagnetic state and the second corresponds to the triangular lattice with $i$-$wave$ altermagnetism. Unless stated otherwise, we set the pairing strength to $J=1.6|t|$.

%%%%%%%%%%%%%%%%%%%%%%%%%%%%%%%%%%%%%%%%%%%%%%%%%%%%%%%%%%%%%%%%%%%%%%%%%%%%%%%%%%%%%%%%%%%%
\subsection*{Square lattice}
\begin{figure*}[ht]
    \centering   
    \includegraphics[scale=0.08]{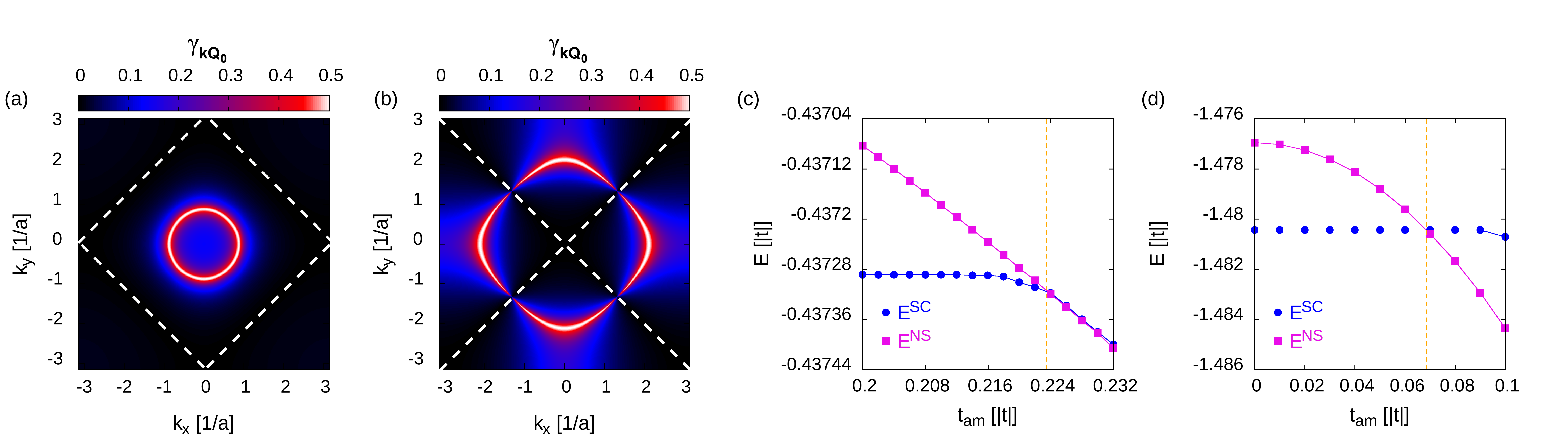} 
     \caption{Cooper pair density $\gamma_{\mathbf{kQ_0}}=\gamma_{\mathbf{kQ_0}}^{\bar{\sigma}\sigma}$, for $\mathbf{Q}_0=(0,0)$, $t_{\mathrm{am}}=0$, and for two selected band fillings $n=0.12$ (a) and $n=0.64$ (b) where the $extended$ $s$-$wave$ and $d$-$wave$ pairing symmetries are stable, respectively. The white dashed lines correspond to the so-called nodal lines where due to symmetry reasons the SC gap has to be zero. In (c) and (d) we show the free energy of the $extended$ $s$-$wave$ (for $n=0.12$) and $d$-$wave$ (for $n=0.64$) superconducting states, respectively, together with the normal state energy, all as functions of $t_{\mathrm{am}}$. The yellow dashed line marks the transition point where SC becomes unstable and $E^{SC}>E^{NS}$.s}
    \label{fig:square_t_j_n}
\end{figure*}

\vspace{2ex}
We start by considering a square lattice system and characterize the influence of altermagnetic phase with $d$-$wave$ symmetry on the stability of superconducting phase in the absence of non-zero momentum Cooper pairing. The $\mathbf{Q}\neq 0$ situation is analyzed in the next step. Similarly as in our previous study, for the case without any altermagnetic ordering, the model with the real-space pairing term shows stability of the spin-singlet superconductivity with $extended$ $s$-$wave$ symmetry of the gap in the low electron concentration regime, as well as the $d$-$wave$ symmetry close to half-filling\cite{Zegrodnik_s_wave}. It should be noted that significant differences appear between the two cases. Namely, as shown by the calculated pair density map in Fig. \ref{fig:square_t_j_n}, for the $extended$ $s$-$wave$ pairing, the Fermi surface is far away from the nodal lines, meaning that a fully gapped situation is realized. On the other hand, for the $d$-$wave$ paired state nodal lines cross the Fermi surface and the gap closses in the diagonal directions. In Figs. \ref{fig:square_t_j_n} (c) and (d) we show the influence of the altemagnetic spin splitting (with increasing $t_{\mathrm{am}}$ parameter) on the pairing for two representative band fillings corresponding to the appearance of the two SC gap symmetries. As can be seen, in both cases, up to some threshold value, the free energy of the obtained superconducting solution is lower than that of the normal state (NS). In that region, the presence of altermagnetism does not change the superconducting gap amplitude, which is $\Delta^{\mathrm{s}}_{\mathrm{ext-s}}=0.068|t|$ and $\Delta^{\mathrm{s}}_{\mathrm{d}}=0.085|t|$ for the $extended$ $s$-$wave$ and $d$-$wave$ paired states, respectively. At some critical spin splitting amplitude the SC state becomes no longer stable since $E^{SC}>E^{NS}$. The reported effect is expected because with increasing $t_{\mathrm{am}}$ the alternating spin splitting (cf. Fig. \ref{fig:theory_1}) is becoming more and more enhanced, leading to Fermi wave vector mismatch, which above some threshold value has a strong destructive influence on the superconducting state. In such a case, less and less $(\mathbf{k}\sigma, -\mathbf{k}\bar{\sigma})$ states can be found around the Fermi surface to form Cooper pairs.

%as shown in more detail in Appendix B. 

In the following we take into account the possibility of non-zero momentum of the Cooper pairs and analyze the general properties of the FF phase induced by the presence of the altermagnetic behavior for the case of low- and high-electron concentration regimes. It should be noted that for a given $n$ the most favorable conditions for the appearance of the FF phase are expected to appear roughly close to the threshold values determined here.

%%%%%%%%%%%%%%%%%%%%%%%%%%%%%%%%%%%%%%%%%%%%%%%%%%%%%%%%%%%%%%%%%%%%%%%%%%%%%%%%%%%%%%%%%%%%%%
\vspace{2ex}
\subsubsection*{Low electron concentration}
\begin{figure*}[ht]
    \centering   
    \includegraphics[scale=0.08]{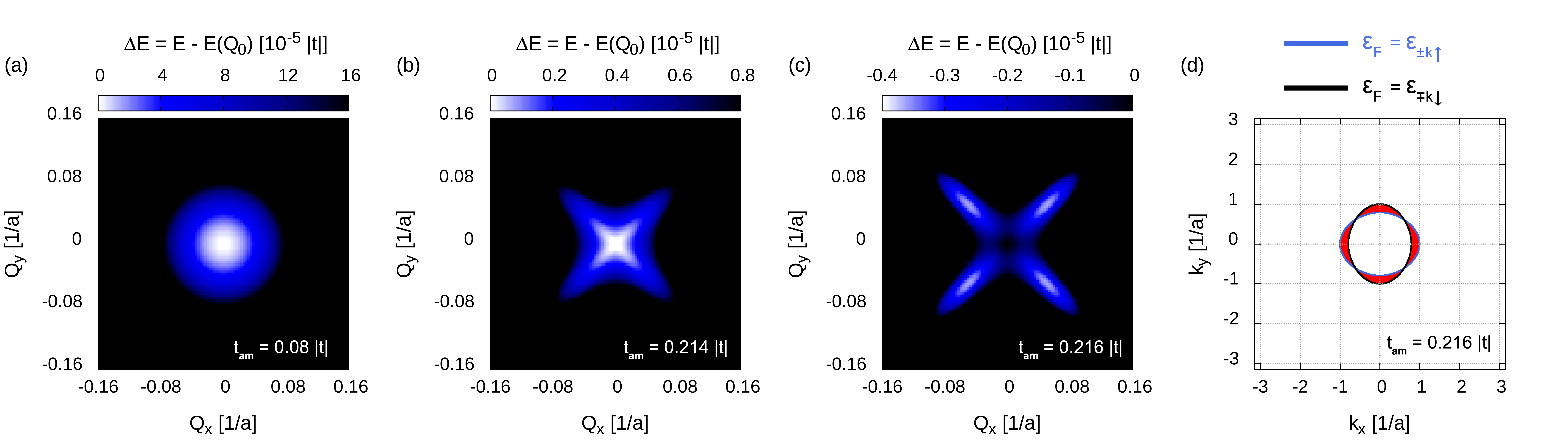}    
     \caption{Free energy of the $extended$ $s$-$wave$ superconducting state as a function of Cooper pair momentum, obtained for $n$ = 0.12. Three different values of the $d$-$wave$ altermagnetic spin splitting $t_{am}$ have been selected: (a) $t_{am}=0.08|t|$; (b) $t_{am}=0.214|t|$ and (c) $t_{am}=0.216|t|$. All energies are shifted relative to energy for $\mathbf{Q}=(0,0)$-paired state. (d) The $\varepsilon_{F}=\varepsilon_{\pm\mathbf{k}\uparrow}$ and $\varepsilon_{F}=\varepsilon_{\mp\mathbf{k}\downarrow}$ Fermi surfaces with four regions of large Fermi wave vector mismatch marked by red color for the $t_{\mathrm{am}}$ value corresponding to the FF phase formation presented in (c).} 
    \label{fig:cooper_square_n_0_12}
\end{figure*}

\vspace{2ex}
In order to study the creation of the FF phase resulting from the interplay between the $extended$ $s$-$wave$ pairing and the $d$-$wave$ altermagnetic spin splitting, we focus on the low electron concentration regime by setting $n=0.12$. In Fig. (\ref{fig:cooper_square_n_0_12}) we show the calculated free energy of the $extended$ $s$-$wave$ superconducting state as a function of Cooper pair momentum for three selected $t_{am}$ values. As one can see, for a relatively small amplitude of the spin splitting, the energy minimum is located at $\mathbf{Q}= (0,0)$, meaning that the standard SC state with zero momentum pairing is stable. However, with increasing $t_{\mathrm{am}}$, noticeable changes in the free energy distribution are reported, with four distinct minima appearing in (c). Those minima correspond to non-zero momentum of the Cooper pairs and are the signatures of the FF phase formation induced by the altermagentic behavior. In Fig. \ref{fig:cooper_square_n_0_12} (d) we show the Fermi surfaces determined by $\varepsilon_{F}=\varepsilon_{\pm\mathbf{k}\uparrow}$ and $\varepsilon_{F}=\varepsilon_{\mp\mathbf{k}\downarrow}$ for the $t_{\mathrm{am}}$ value corresponding to the formation of the FF phase presented in (c). The regions where the determined Fermi surfaces are far from each other are marked by red and correspond to significant Fermi wave vector mismatch which is detrimental to the pairing. The non-zero center-of-mass momentum of the Cooper pairs allows to reduce the Fermi wave vector mismatch significantly.

\begin{figure*}[h]
    \centering   
    \includegraphics[scale=0.32]{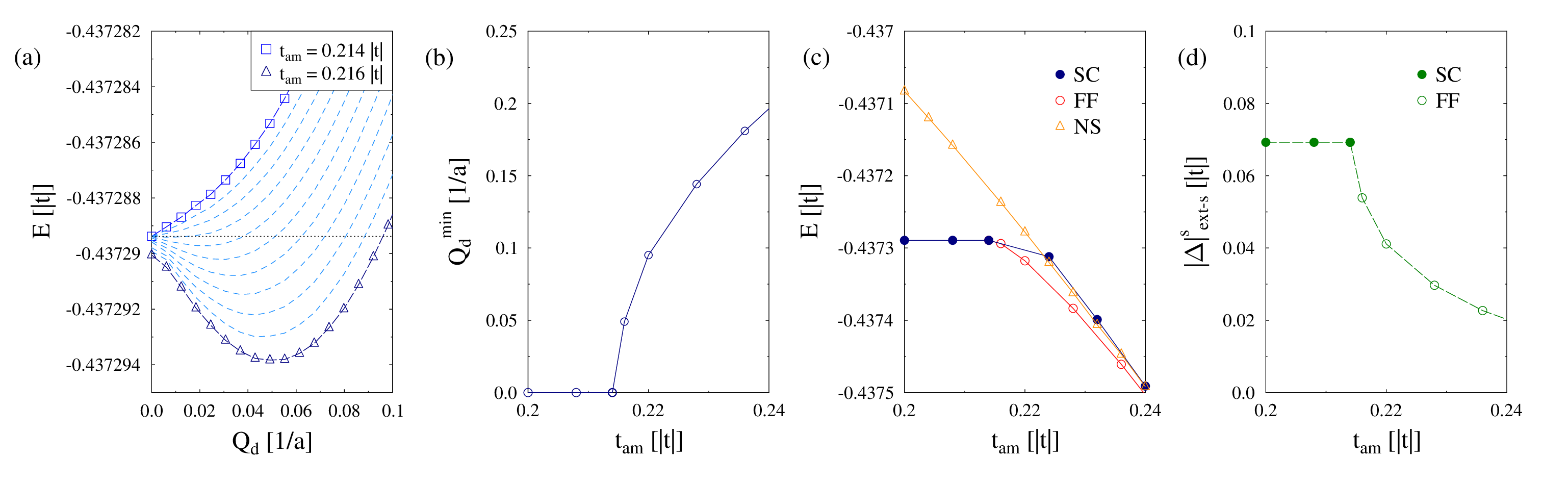} 
    \caption{(a) Free energy of the $extended$ $s$-$wave$ superconducting state as a function of $Q_d=Q_x=Q_y$, where the Cooper pair momentum has the form $\mathbf{Q}=(Q_x,Q_y)$. Dashed lines represents curves calculated for $t_{am}$ from range $0.214|t|$-$0.216|t|$, with step of $0.0002|t|$. (b) Values of $Q_d$ corresponding to the Cooper pair momentum minimizing free energy, as function of altermagnetic spin splitting amplitude. (c) Free energy of the SC (paired state with $\mathbf{Q}=0$), FF (paired state with $\mathbf{Q}\neq0$), and NS (normal state) as a function of $t_{\mathrm{am}}$. (d) Superconducting gap amplitudes versus $t_{am}$ close to the transition point between the SC and FF states.}
    \label{fig:q_min_square_n_0_12}
\end{figure*}

To analyze in more detail the transition to the FF phase with increasing altermagnetic spin splitting amplitude, in Fig. \ref{fig:q_min_square_n_0_12} (a) we plot the free energy of the superconducting state as a function of $Q_d=Q_x=Q_y$ for eleven different $t_{\mathrm{am}}$ values. As one can see with increasing $t_{\mathrm{am}}$ a free energy minimum for non-zero $Q_d$ is starting to form gradually. The critical value of $t_{am}$ for the creation of the FF state corresponds to the lowest $t_{am}$ for which the energy minimum appears at $Q_d\neq 0$. Note that by increasing $Q_d>0$ we move along the diagonal direction in the Cooper pair momentum space, crossing one of the minima visible in Fig. \ref{fig:cooper_square_n_0_12} (c). Based on the data obtained, we have extracted the $Q_d$ values which lead to the minimization of the free energy for different $t_{\mathrm{am}}$. The result is presented in Fig. \ref{fig:q_min_square_n_0_12} (b), where one can see that the paired phase with non-zero Cooper pair momentum becomes more energetically favorable above $t_{\mathrm{am}} \approx 0.215|t|$, which marks the transition to the FF phase.

Moreover, the location of the energy minimum along the diagonal $Q_d$ axis changes with increasing $t_{am}$. It is due to the fact that, increasing altermagnetic spin splitting amplitude generates larger Fermi wave vector mismatch, hence larger values of the Cooper pair momenta are also necessary to compensate for it. For the sake of completeness in (c) we show the calculated free energies of the SC, FF and NS states. As one can see the FF solution  allows for the pairing to survive above the threshold value where the zero momentum pairing is already unstable due the significant Fermi wave vector mismatch. In Fig. \ref{fig:q_min_square_n_0_12} (d), we show how the superconducting gap parameter changes close to the transition point between the SC and FF states.

\begin{figure*}[h]
    \centering   
         \includegraphics[scale=0.08]{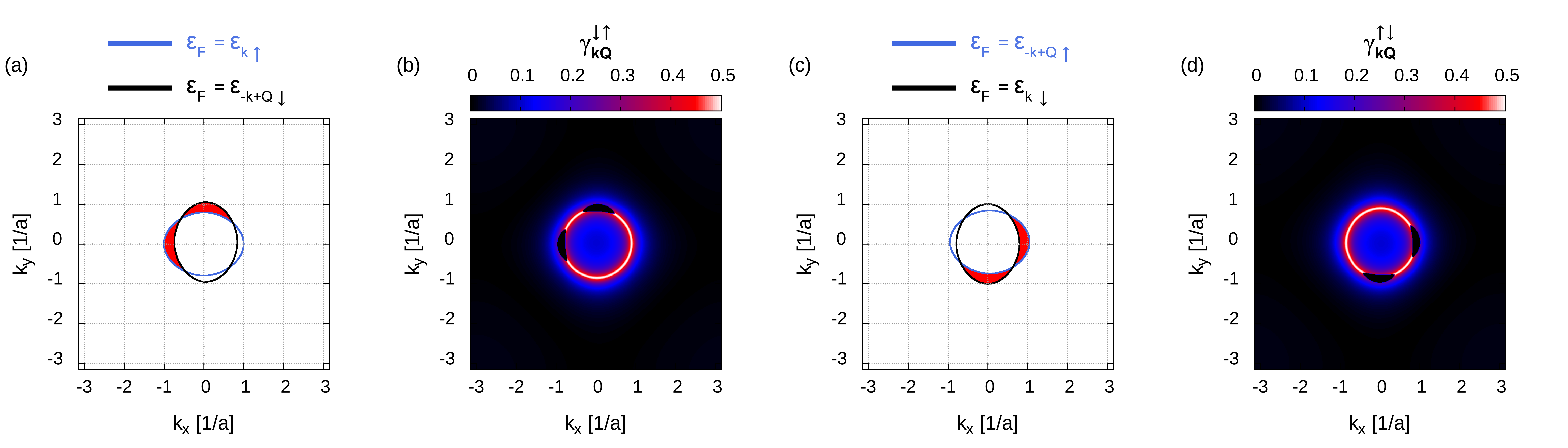}
    \caption{The $\varepsilon_{\mathbf{k}\sigma}$ and $\varepsilon_{\mathbf{-k+Q}\bar{\sigma}}$ Fermi surfaces [(a) and (c)] between which the pairing appears for the case of $d$-$wave$ altermagnetic spin splitting amplitude $t_{\mathrm{am}}=0.216|t|$, for which the FF phase is stable with non-zero centre-of-mass momentum of the Cooper pairs $\mathbf{Q}\approx(0.05,0.05)\;1/a$ (cf. Figs. \ref{fig:cooper_square_n_0_12}). The red color marks the regions in which significant Fermi wave vector mismach appears. The pair density $\gamma^{\bar{\sigma}\sigma}_{\mathbf{kQ}}$ for the same model parameters is shown in (b) and (d). Note that the depairing regions for which $\gamma^{\bar{\sigma}\sigma}_{\mathbf{kQ}}=0$ in (b) and (d) correspond to the areas of significant Fermi wave vector mismatch seen in (a) and (c). Due to the non-zero value of $\mathbf{Q}$ pairing takes place at a large extent of the Fermi surface.}
    \label{fig:pair_dens_square_n_0_12}
\end{figure*}

In order to gain some more insight into the mechanism of FF phase formation in the presence of altemagnetic spin splitting, we analyze the relative position of the spin up and down Fermi surfaces as well as the calculated Cooper pair density $\gamma^{\bar{\sigma}\sigma}_{\mathbf{Qk}}$. As shown in Fig. \ref{fig:cooper_square_n_0_12} (d) in the considered case there are four regions in $\mathbf{k}$-space in which a significant Fermi wave vector mismatch between $\varepsilon_{F}=\varepsilon_{\pm\mathbf{k}\uparrow}$ and $\varepsilon_{F}=\varepsilon_{\mp\mathbf{k}\downarrow}$ Fermi surfaces appears. Fig. \ref{fig:pair_dens_square_n_0_12} corresponds to situation in which the FF phase is already stable with $\mathbf{Q}\approx(0.05,0.05)\;1/a$\;. In (a) and (c) we present the Fermi surfaces between which the pairing appears in the FF phase taking into account the non-zero center-of-mass momentum of the Cooper pairs $\mathbf{Q}$ which shifts the spin $\sigma$ Fermi surface with respect to the spin $\bar{\sigma}$. As a result of the non-zero momentum pairing, instead of four regions with significant Fermi wave vector mismatch, there are only two marked in red. This allows for the pairing to appear at the large extent of the Fermi surfaces where non-zero pair-density is visible in (b) and (d). Nevetheless, a two depairing regions in $\mathbf{k}$-space still appear where $\gamma^{\bar{\sigma}\sigma}_{\mathbf{Qk}}=0$.

It should be noted that in the considered case of $extended$ $s$-$wave$ paired state which is stable in the low-electron-concentration regime, the SC gap is practically isotropic. Therefore, the orientation of the Cooper pair momentum in the FF state is completely determined by the symmetry of the altermagnetic order parameter. As we could see for the case of $d$-$wave$ altermagnet the $\mathbf{Q}$-vectors are aligned with the diagonal directions for which the spin splitting vanishes. This leads to four energy minima located at the $k_x=k_y$ and $k_x=-k_y$ directions (cf. Fig. \ref{fig:cooper_square_n_0_12}). For the $g$-$wave$ symmetry which is not considered here in detail, there would be additional directions corresponding to vanishing spin-splitting\;\cite{YuJun_Bilayer_AM,Li2025,Beida2025}. Namely, apart from the diagonal directions, also additional ones appear which correspond to $k_x=0$ and $k_y=0$. Therefore, in analogy to the case considered above, one can expect that for the $g$-$wave$ altermagneti there should be eight $\mathbf{Q}$-vectors of the FF state lying at the $k_x=0$, $k_y=0$, $k_x=k_y$, and $k_x=-k_y$ directions (two minima per each direction). Nevertheless, a complete detailed analysis of the interplay between $g$-$wave$ altermagnet with various types of SC symmetry factors is beyond the scope of this paper. We should see progress along this line soon.

%%%%%%%%%%%%%%%%%%%%%%%%%%%%%%%%%%%%%%%%%%%%%%%%%%%%%%%%%%%%%%%%%%%%%%%%%%%%%%%%%%%%%%%%%%%%%%

\vspace{1ex}
\subsubsection*{Influence of external magnetic field}
\vspace{1ex}
For the sake of completeness, we supplement our analysis in the low electron concentration regime with the situation corresponding to non-zero magnetic field oriented in the $z$ direction. It should be noted that the spin splitting generated by external magnetic field is isotropic; therefore, in terms of the symmetry classification, non-zero values of $B_z$ introduce an $s$-$wave$ spin splitting. Hence, for the case of square lattice with both $t_{\mathrm{am}}\neq 0$ and $B_z\neq 0$ we actually consider a mixed $d$-$wave$ and $s$-$wave$ spin splitting in the system.
\begin{figure}[h!]
    \centering   
            \includegraphics[scale=0.08]{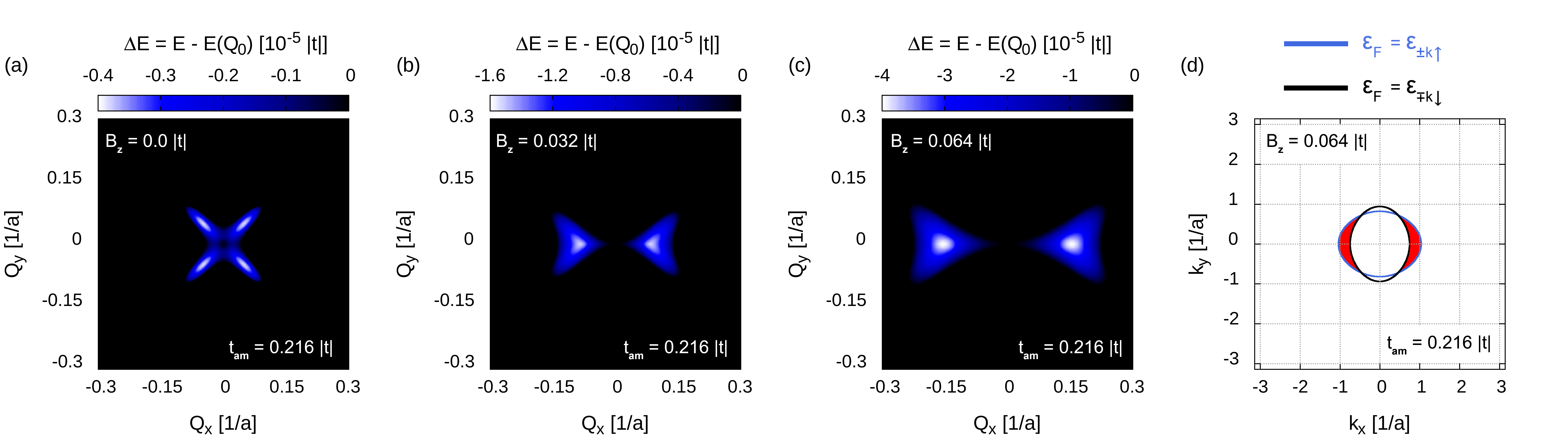}  
    \caption{Free energy as a function of Cooper pair momentum for $t_{\mathrm{am}}=0.216|t|$ for selected values of magnetic field oriented along the $z$-direction: (a) $B_z=0.0$, (b) $B_z=0.032|t|$, and (c) $B_z=0.064|t|$. All energy values were shifted relative to energy obtained for the $\mathbf{Q}=(0,0)$ case. (d) the $\varepsilon_{F}=\varepsilon_{\pm\mathbf{k}\uparrow}$ and $\varepsilon_{F}=\varepsilon_{\mp\mathbf{k}\downarrow}$ Fermi surfaces with two regions of large Fermi wave vector mismatch marked by red color for the $t_{\mathrm{am}}$ value corresponding to the FF phase formation presented in (c).}
    \label{fig:magnetic_cooper}
\end{figure}

\begin{figure}[ht!]
    \centering   
        \includegraphics[scale=0.08]{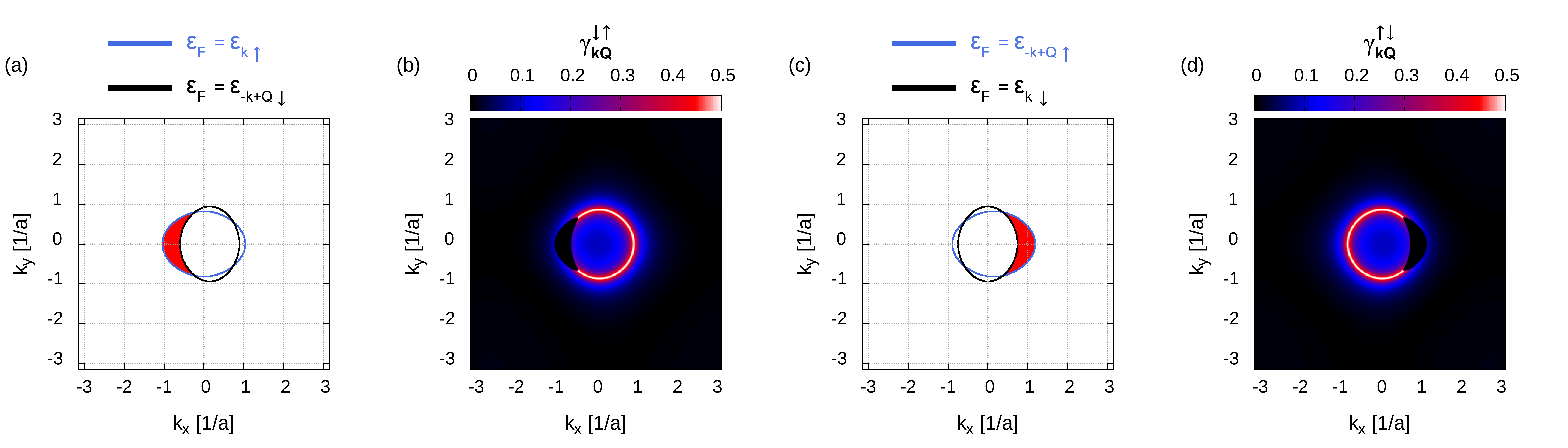}        
    \caption{The $\varepsilon_{\mathbf{k}\sigma}$ and $\varepsilon_{\mathbf{-k+Q}\bar{\sigma}}$ Fermi surfaces [(a) and (c)] between which the pairing appears for the case of $d$-$wave$ altermagnetic spin splitting amplitude $t_{\mathrm{am}}=0.216|t|$ and $B_z=0.064|t|$, for which the FF phase is stable with non-zero centre-of-mass momentum of the Cooper pairs $\mathbf{Q}\approx(0.15,0.0)\;1/a$ (cf. Fig. \ref{fig:magnetic_cooper}) The red area marks the regions in which significant Fermi wave vector mismach appears. The pair density $\gamma^{\bar{\sigma}\sigma}_{\mathbf{Qk}}$ for the same model parameters is shown in (b) and (d). Note that the depairing regions for which $\gamma^{\bar{\sigma}\sigma}_{\mathbf{Qk}}=0$ in (b) and (d) correspond to the areas of significant Fermi wave vector mismatch seen in (a) and (c). Due to the non-zero value of $\mathbf{Q}$, pairing takes place at a large extent of the Fermi surface.}
    \label{fig:square_n_0_12_pair_density_magn}
\end{figure}

In Fig. \ref{fig:magnetic_cooper} we show the free energy of the $extended$ $s$-$wave$ superconducting state as a function of the Cooper pair momentum for selected values of the external magnetic field. As one can see with increasing $B_z$ the $\mathbf{Q}$ vectors corresponding to the FF state move towards the $Q_x$-axis forming two free energy minima. This is because for $B_z>0$ the spin down Fermi surface shrinks with respect to the spin up correspondent. In such case instead of four regions of significant Fermi wave vector mismatch [cf Fig. \ref{fig:cooper_square_n_0_12} (d)] we have only two, which cross symmetrically the $k_x$-axis [cf. Fig. \ref{fig:magnetic_cooper}(d)]. Therefore, it is most convenient for the $\mathbf{Q}$ vector to also be directed along the horizontal axis, which significantly reduces one of the areas with a significant Fermi wave vector mismatch. This allows for only one depairing region appearing in the $\mathbf{k}$-space, with Cooper pairs created along a significant extent of the Fermi surface, as shown in Fig. \ref{fig:square_n_0_12_pair_density_magn}.

\begin{figure*}[h!]
    \centering   
    \includegraphics[scale=0.3]{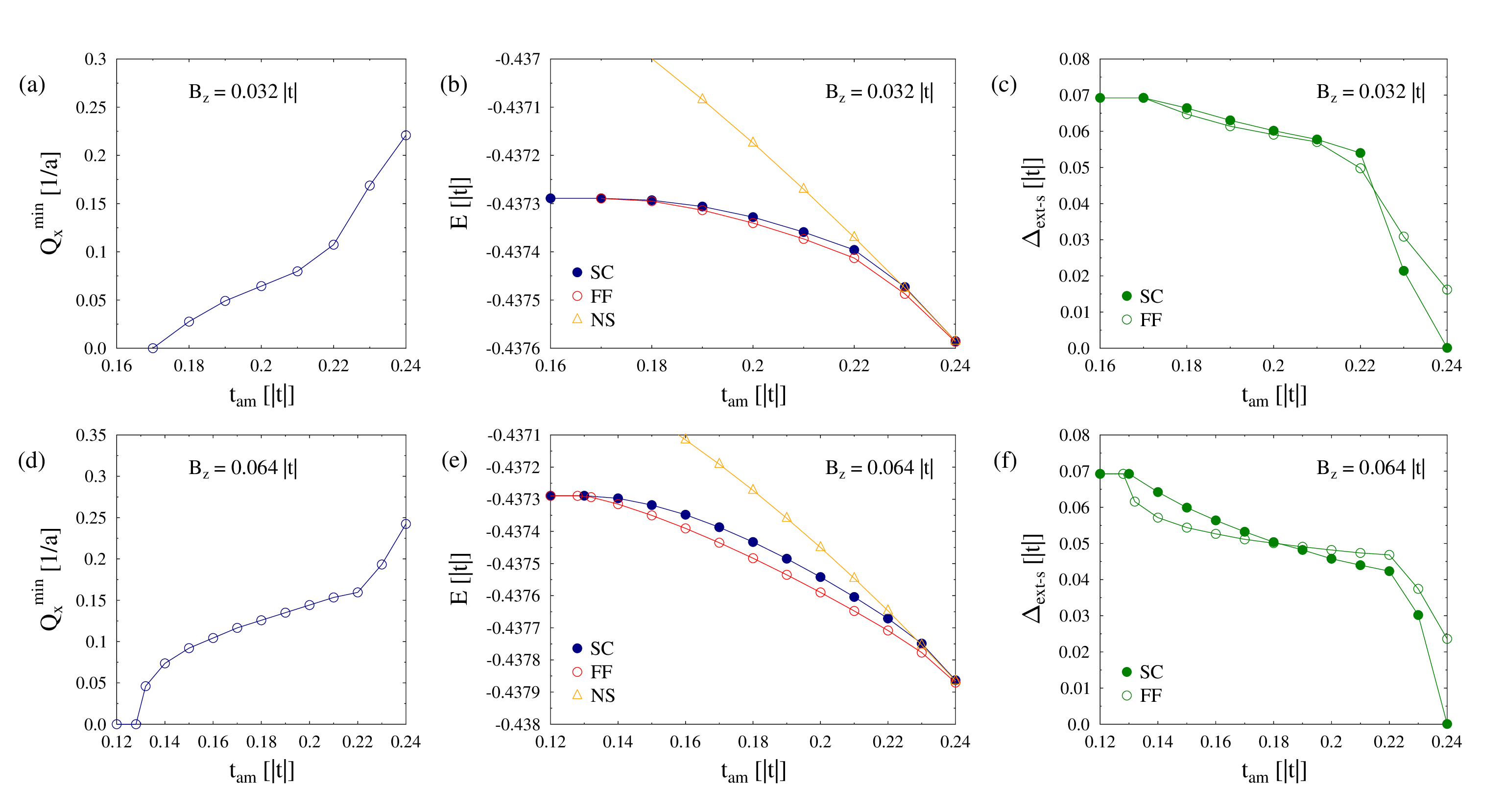} 
    \caption{Cooper pair momentum in the $Q_x$-direction [$\mathbf{Q}=(Q_x,0)$] which minimizes the free energy and corresponds to the stability of the FF phase as a function of the altermagnetic spin splitting amplitude for two selected values of the magnetic field $B_z=0.032|t|$ (a) and $B_z=0.064|t|$ (b). In (b) and (e) we show the free energy of the SC, FF and NS states for the same model parameters as in (a) and (d), respectively. Note the stability of the FF phase. In (c) and (f) we show the superconducting gap amplitude corresponding to the FF and SC phase for the same model parameters as in (a) and (d), respectively. }
    \label{fig:square_magnetic_n_0_12_q_min}
\end{figure*}

Similarly as before, the larger the altermagnetic spin splitting is, the larger also the Cooper pair momentum corresponding to the stability of the FF phase, as shown in Fig. \ref{fig:square_magnetic_n_0_12_q_min}. In (b) and (e) we show that in the presence of external magnetic field the stability range of the FF phase in terms of the $t_{\mathrm{am}}$ amplitude is much wider than for the $B_z=0$ case. This is because both magnetic field and altermagnetism generate spin-splitting which enhances the Fermi wave vector mismatch. Large enough Fermi wave vector mismatch generates non-zero momentum pairing. Therefore, if the two factors appear simultaneously, then the FF phase is more likely to appear. It should be noted that in the case of both $t_{\mathrm{am}}\neq0$ and $B_z\neq0$ we have obtained a small admixture of a $d$-$wave$ pairing symmetry to the dominant $extended$-$s$-$wave$ symmetry. However, the former is two orders of magnitude smaller than the latter, hence it should be considered as a negligible effect.

There is an interesting aspect related with the interplay between the external magnetic field and altermagnetism in the context of the FF phase formation. Namely, as we shown for $B_z>0$ the Cooper pair momentum is directed along the $x$-axis. However, for the case of $B_z<0$ the spin up Fermi surface shrinks with respect to the spin down correspondent, meaning that the depairing regions are going to cross symetrically to $k_y$ axis instead of $k_x$. This would create an FF phase with Cooper pair momentum $\mathbf{Q}=(0, \pm Q_y^{\mathrm{min}})$, which means that by changing the orientation of the magnetic field, one can change the direction of the Cooper pair momentum by $\pi/2$. Such effect can be potentially interesting when it comes to various applications. Namely, in the context of highly tunable direction-dependent superconducting diode. However, this would require the presence of spin-orbit interaction\cite{Fukaya_2025,Yusuke_2025}, which is beyond the scope of this study.

%%%%%%%%%%%%%%%%%%%%%%%%%%%%%%%%%%%%%%%%%%%%%%%%%%%%%%%%%%%%%%%%%%%%%%%%%%%%%%%%%%%%%%%%%%%%%
\vspace{2ex}
\subsubsection*{High electron concentration}
\vspace{2ex}

\begin{figure}[ht]
    \centering   
   \includegraphics[scale=0.08]{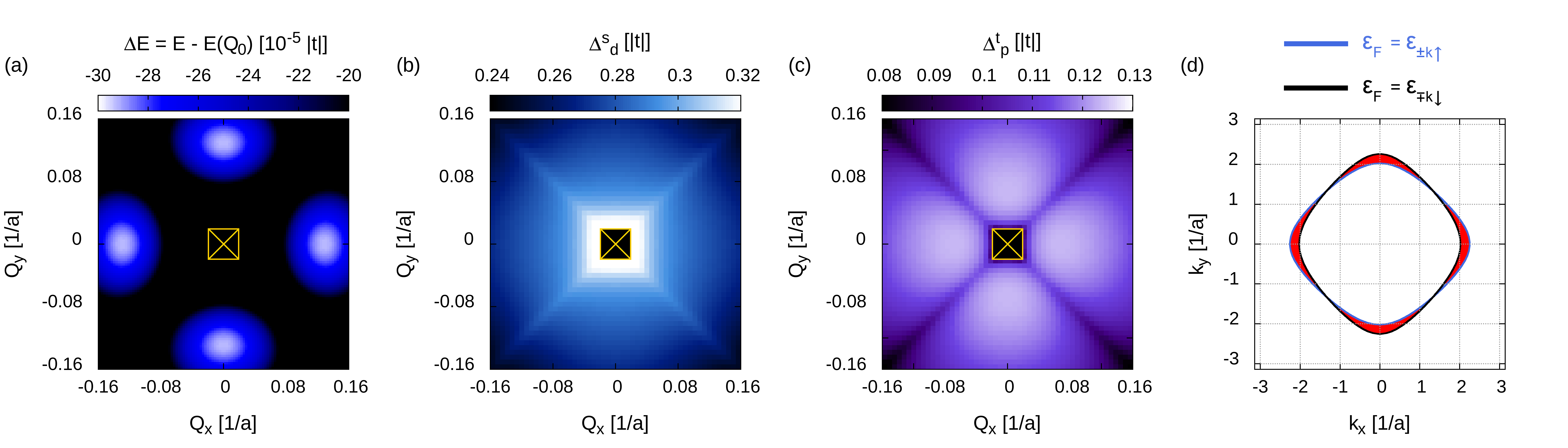} 
     \caption{(a) Free energy of the superconducting solution as a function of non-zero Cooper pair momentum calculated for $t_{am}$ = 0.064 $|t|$. All free energy values are shifted relative to free energy of state with zero momentum of Cooper pair with pure $d$-wave symmetry. Yellow square at center shows area of unstable solution with mixed $p-d$ pairing. Only pure $d$-$wave$ solution was found there. (b) Superconducting gap amplitude $\Delta^s_d$ in mixed pairing state as a function of $\mathbf{Q}$. (c) Superconducting gap amplitude $\Delta^t_p$ in mixed pairing state as a function of $\mathbf{Q}$. (d) the $\varepsilon_{F}=\varepsilon_{\pm\mathbf{k}\uparrow}$ and $\varepsilon_{F}=\varepsilon_{\mp\mathbf{k}\downarrow}$ Fermi surfaces with four regions of large Fermi wave vector mismatch marked by red color for the $t_{\mathrm{am}}$ value corresponding to the FF phase formation presented in (a).}
    \label{fig:cooper_square_n_0_64}
\end{figure}

Here we consider a substantially higher electron concentration ($n=0.64$) which corresponds to the stability of the $d$-$wave$ supercodncuting state. As already shown in Fig. \ref{fig:square_t_j_n} (d) for the selected band filling and without the inclusion of non-zero momentum pairing the pure $d$-$wave$ SC phase survives up to $t_{\mathrm{am}}=0.07|t|$. In Fig. \ref{fig:cooper_square_n_0_64} we show the free energy of the paired state as a function of Cooper pair momentum for $t_{\mathrm{am}}=0.064|t|$ where four distinct minima appear which correspond to the stability of the FF state similarly to the case of $extended$ $s$-$wave$ superconductivity. Nevertheless, there are two main differences with respect to the previous considerations. Namely, now the FF phase is formed for the Cooper pair momenta oriented along the $Q_x$- and $Q_y$-axes instead of the diagonal directions in the $\mathbf{Q}$-plane (cf. Fig. \ref{fig:cooper_square_n_0_12}). Furthermore, as visualized in Figs. \ref{fig:cooper_square_n_0_64} (b) and (c), this time the FF paired state corresponds to a mixture of $d$-$wave$ (singlet) and $p$-$wave$ (triplet) pairing amplitudes. It should be noted that the two symmetries do not correspond to two separate superconducting solutions but a single solution with both pairing amplitudes being nonzero.

\begin{figure}[h!]
    \centering   
         \includegraphics[scale=0.32]{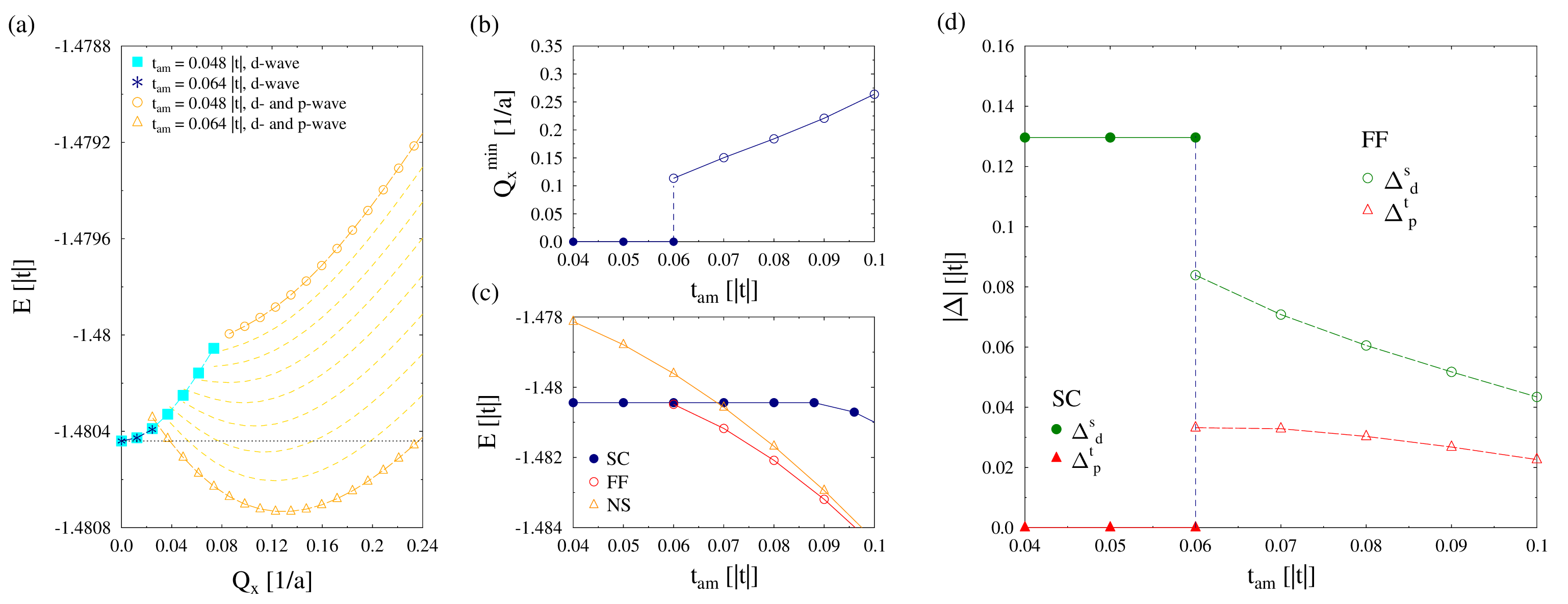}
    \caption{(a) Free energy as a function of Cooper pair momentum $Q_x$ [for $\mathbf{Q}=(Q_x,0$)] calculated for different $t_{am}$ values. Dashed lines points represents curves calculated for $t_{am}$ from 0.048 $|t|$ to 0.064 $|t|$, with step of 0.004 $|t|$. (b) Cooper pair momentum minimizing the free energy as a function of $t_{\mathrm{am}}$. (c) Free energy of the SC, FF, and NS states a functions of $t_{\mathrm{am}}$. (d) Superconducting gap amplitudes versus $t_{am}$ for SC and FF states close to the transition point.}
    \label{fig:q_min_square_n_0_64}
\end{figure}

\begin{figure}[h!]
    \centering   
\includegraphics[scale=0.08]{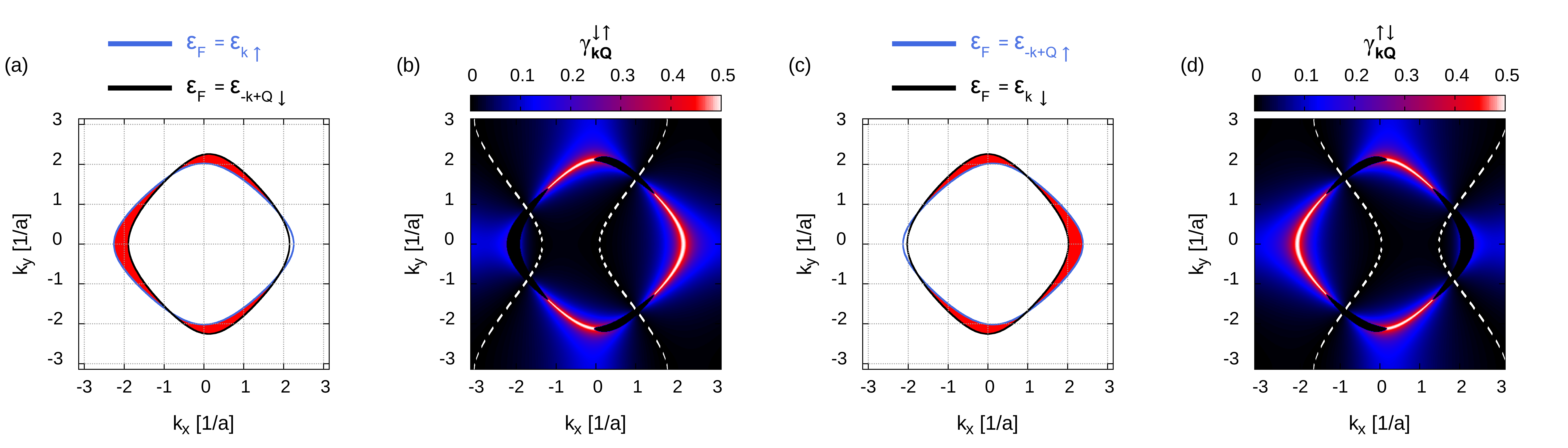}
    \caption{The $\varepsilon_{\mathbf{k}\sigma}$ and $\varepsilon_{\mathbf{-k+Q}\bar{\sigma}}$ Fermi surfaces [(a) and (c)] between which the pairing appears for the case of $d$-$wave$ altermagnetic spin splitting amplitude $t_{\mathrm{am}}=0.064|t|$, for which the FF phase is stable with non-zero centre-of-mass momentum of the Cooper pairs $\mathbf{Q}\approx(0.15,0.0)\;1/a$ (cf. Fig. \ref{fig:cooper_square_n_0_64}). The red area marks the regions in which significant Fermi wave vector mismach appears. The pair density $\gamma^{\bar{\sigma}\sigma}_{\mathbf{Qk}}$ for the same model parameters is shown (b) and (d). Note that the depairing regions for which $\gamma^{\bar{\sigma}\sigma}_{\mathbf{Qk}}=0$ in (b) and (d) correspond to the areas of significant Fermi wave vector mismatch seen in (a) and (c). Due to the non-zero value of $\mathbf{Q}$, pairing takes place at a large extent of the Fermi surface. White dashed lines correspond to nodal lines of the resulting pairing symmetry.}
    \label{fig:pair_dens_square_n_0_64}
\end{figure}

To analyze the formation of the FF phase with gradual increase in altermagnetic spin splitting, we carried out calculations along the $Q_x$-axis of the Cooper pair momentum for a set of $t_{\mathrm{am}}$ values. As one can see in Fig. \ref{fig:q_min_square_n_0_64} (a), a free energy minimum corresponding to $Q_x\neq0$ is beginning to form with gradually increasing $t_{\mathrm{am}}$ values. For $Q_x\gtrsim0$ we have obtained a pure $d$-$wave$ SC solution marked by blue full points. However, for larger values of $Q_x$ a mixed $d$-$wave$/$p$-$wave$ solution was found which is marked by orange empty points. With increasing altermagnetic amplitude, the range of $Q_x$ values corresponding to pure $d$-$wave$ paired states gets narrower and narrower. Therefore, the stable FF solution has a mixed $d$-$wave$/$p$-$wave$ symmetry of the supercondcuting gap. The $t_{\mathrm{am}}$ dependence of the Cooper pair momentum $Q^{\mathrm{min}}_x$ corresponding to the minimum free energy is provided in Fig. \ref{fig:q_min_square_n_0_64} (b), where the transition to the mixed $d/p$-$wave$ paired FF state coincides with a discontinuous jump of $Q^{\mathrm{min}}_x$. In Fig. \ref{fig:q_min_square_n_0_64} (c), we show that the free energy of the FF state is below that corresponding to both the SC and NS states beginning from some critical $t_{\mathrm{am}}$. As one can see in Fig. \ref{fig:q_min_square_n_0_64} (d) at the transition from SC to FF an additional $p$-wave pairing symmetry sets in, which results in about 30\% decrease of $d$-$wave$ superconducting gap amplitude.  However, the mixed FF state can survive up to larger $t_{\mathrm{am}}$ values where trivial $d$-$wave$ solution is already unstable as shown in (c).

It should be noted that the two factors that determine the orientation of the Cooper pair momentum in the FF phase, are: (i) the distribution of the areas in $\mathbf{k}$-space which correspond to a significant wave vector mismatch between the spin up and down Fermi surfaces; (ii) the symmetry of the SC order parameter which determines the distribution of the gap across the Fermi surface. For the case of $extended$ $s$-$wave$ pairing we have a fully gapped (isotropic) situation, therefore the orientation of the Cooper pair momentum is completely determined by (i). This leads to $\mathbf{Q}$ oriented along the diagonal directions, which significantly decreases the wave vector mismatch for the two out of the four regions shown in Fig. \ref{fig:cooper_square_n_0_12} (d). For the case of the $d$-$wave$ paired state, the superconducting gap closes at the diagonal directions due to the appearance of the so-called nodal lines. Therefore, to minimize the Fermi wave vector mismatch and maximize the total SC gap at the Fermi surface the Cooper pair momenta along the vertical and horizontal directions are most suitable. Additionally, a $p$-$wave$ admixture appears to the pairing symmetry, which allows to move slightly the nodal lines and optimize even more the SC gap distribution across the Fermi surfaces. The new nodal lines modified by the admixture of the $p$-$wave$ symmetry are marked by a white dashed line in Fig. \ref{fig:pair_dens_square_n_0_64}. As one can see they are moved away from the regions in which the Fermi wave vector mismatch is small allowing the gap to open there. It should be noted that the $p$-$wave$ symmetry which allows to modify the arrangement of the nodal lines is of odd symmetry; therefore, it is compatible with the spin-triplet state of the Cooper pairs. As a result, the obtained FF state is of mixed singlet-triplet character.

%%%%%%%%%%%%%%%%%%%%%%%%%%%%%%%%%%%%%%%%%%%%%%%%%%%%%%%%%%%%%%%%%%%%%%%%%%%%%%%%%

\section*{Triangular lattice}

To study the interplay between superconductivity and $i$-$wave$ altermagnetism in a triangular lattice, we choose $n=0.64$ for which the fully gapped $d+id$ superconducting state is realized. Similarly as in previous cases, also here the altermagnetic spin splitting has a negative influence on the paired state. However, for $J=1.6|t|$ the threshold value for the destruction of superconductivity is very small, which leads to numerical complications in the analysis of non-zero momentum pairing. Therefore, for convenience, we set $J=3.2|t|$, which allows us to increase the range of SC stability and carry out the calculations efficiently in reasonable time.

\begin{figure}[h!]
    \centering   
    \includegraphics[scale=0.08]{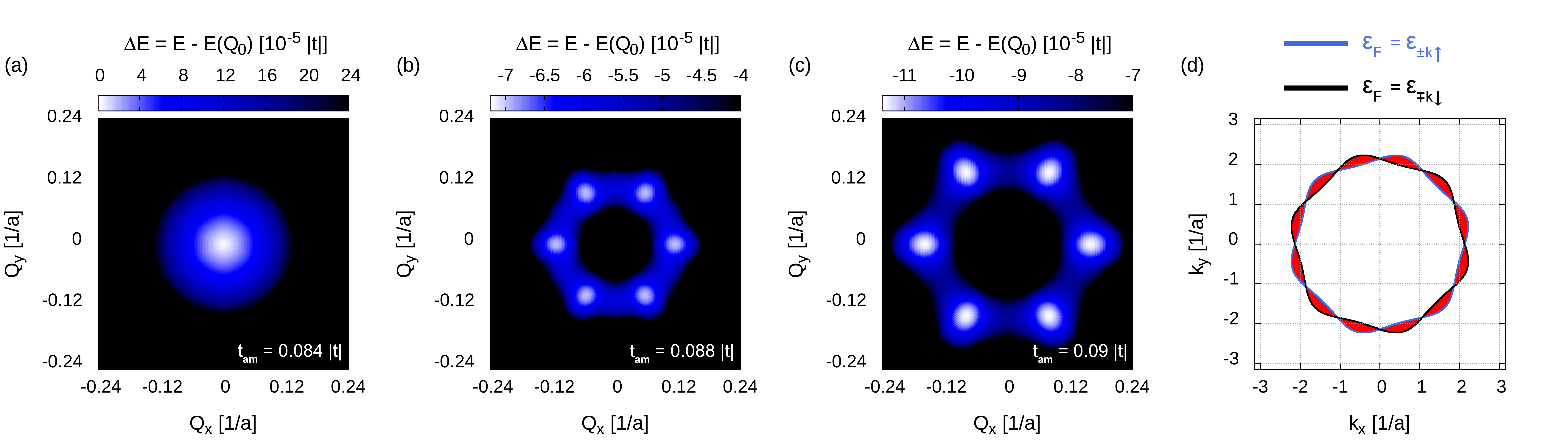}    
     \caption{Free energy as a function of Cooper pair momentum $\mathbf{Q}$ calculated for selected values of the altermagnetic spin splitting amplitude (a) $t_{am}=0.084|t|$, (b) $t_{am}=0.088|t|$ and (c) $t_{am}=0.09|t|$. Symbol $E(Q_0)$ stands for free energy of state with zero Cooper pair momentum. (d) the $\varepsilon_{F}=\varepsilon_{\pm\mathbf{k}\uparrow}$ and $\varepsilon_{F}=\varepsilon_{\mp\mathbf{k}\downarrow}$ Fermi surfaces with twelve regions of large Fermi wave vector mismatch marked by red color for the $t_{\mathrm{am}}$ value corresponding to the FF phase formation presented in (c).}
    \label{fig:triangular_cooper}
\end{figure}

Similarly as in the previously considered cases, also here we report on the stability of FF phase induced by the altermagnetic spin splitting. In Fig. \ref{fig:triangular_cooper} we show the free energy minima for non-zero Cooper pair momentum appearing for large enough values of the altermagnetic spin splitting amplitude. In contrast to the previous situations, here six minima appear that move away from the $(0,0)$ point with increasing $t_{\mathrm{am}}$. The orientation of the $\mathbf{Q}$ vectors which correspond to the stability of the FF state fulfill the $C_6$ symmetry of the triangular lattice with $i$-$wave$ altermagnetic ordering.

\begin{figure}[h!]
    \centering   
\includegraphics[scale=0.32]{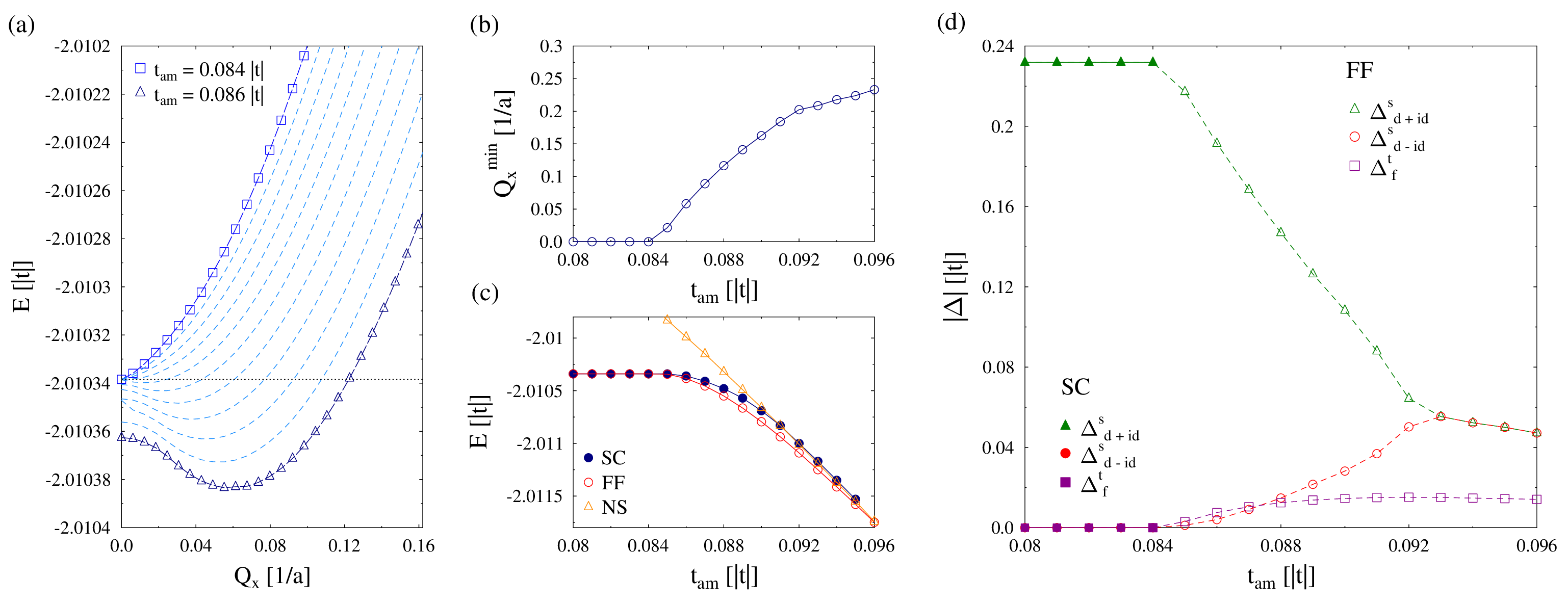} 
       \caption{(a) Free energy as a function of Cooper pair momentum $Q_x$ [for $\mathbf{Q}=(Q_x,0)$] calculated for different $t_{\mathrm{am}}$ values. Dashed lines correspond to $t_{am}$ between 0.084 $|t|$ and 0.086 $|t|$, with step of 0.0002 $|t|$. (b) Cooper pair momentum minimizing the free energy as a function of $t_{\mathrm{am}}$. (c) Free energy of the SC, FF, and NS states as functions of $t_{\mathrm{am}}$. (d) Superconducting gap amplitudes versus $t_{\mathrm{am}}$ for the SC and FF states.}
    \label{fig:q_min_triangle}
\end{figure}

In Fig. \ref{fig:q_min_triangle}(a) we show the gradual formation of the FF state with $\mathbf{Q}=(Q_x,0)$ by increasing $t_{\mathrm{am}}$. The extracted Cooper pair momentum $Q_x^{\mathrm{min}}$ that minimizes the free energy as a function of $t_{\mathrm{am}}$ is provided in (b). As one can see in (c) the non-zero momentum pairing allows to lower down the free energy of the paired state as the $t_{\mathrm{am}}$ is incresed. Note that in (b) a transition point is visible at which the dependence $Q_x^{\mathrm{min}}(t_{\mathrm{am}})$ changes its character. To analyze the origin of such behavior, we focus on the superconducting gap amplitude of the FF phase as a function of $t_{\mathrm{am}}$ shown in (d). As one can see, the obtained FF phase is of mixed symmetry and the mentioned transition point corresponds to change in the composition of particular symmetry components. Namely, before the transition the FF phase corresponds to $d+id$ symmetry with some small (but increasing with $t_{\mathrm{am}}$) admixture of $d-id$ pairing together with a spin-triplet $f$-$wave$ component. After the transition, we obtain $\Delta^s_{d+id}=\Delta^s_{d-id}$, meaning that only the real part of the $d\pm id$ pairing survives, leading to purely real $d$-$wave$ symmetry mixed with an $f$-$wave$ only. Nevertheless, other superconducting gaps ($exended$ $s$-$wave$ and $p\pm ip$) also appear here, however, those are two orders of magnitude smaller, so should be considered as negligible.

\begin{figure}[h!]
    \centering   
        \includegraphics[scale=0.08]{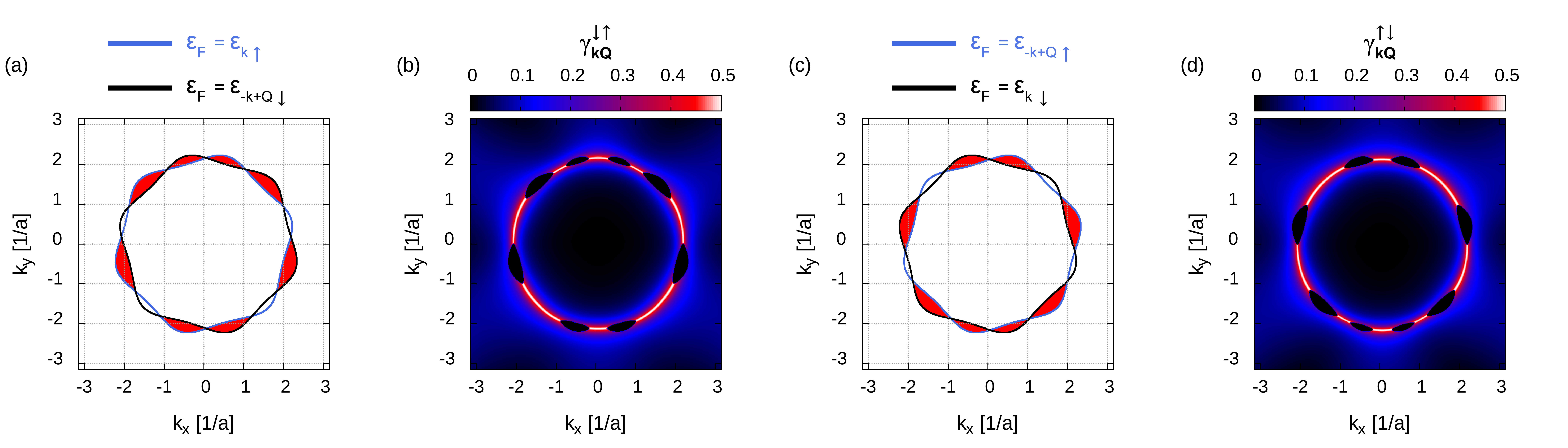}
    \caption{The $\varepsilon_{\mathbf{k}\sigma}$ and $\varepsilon_{\mathbf{-k+Q}\bar{\sigma}}$ Fermi surfaces [(a) and (c)] between which the pairing appears for the case of $i$-$wave$ altermagnetic spin splitting amplitude $t_{\mathrm{am}}=0.09|t|$, for which the FF phase is stable with non-zero centre-of-mass momentum of the Cooper pairs $\mathbf{Q}\approx(0.18,0.0)\;1/a$ (cf. Fig. \ref{fig:triangular_cooper}). The red area marks the regions in which significant Fermi wave vector mismach appears. The pair density $\gamma^{\bar{\sigma}\sigma}_{\mathbf{Qk}}$ for the same model parameters is shown (b) and (d). Note that the depairing regions for which $\gamma^{\bar{\sigma}\sigma}_{\mathbf{Qk}}=0$ in (b) and (d) correspond to the areas of significant Fermi wave vector mismatch seen in (a) and (c). Due to the non-zero value of $\mathbf{Q}$, pairing takes place at a large extent of the Fermi surface.}
    \label{fig:triangular_pair_density}
\end{figure}

For the sake of completeness, we analyze the pair density in the $\mathbf{k}$-space as well as the spin-up and spin-down Fermi surfaces shifted by the Cooper pair momentum, which corresponds to the stability of the FF phase for $t_{\mathrm{am}}=0.09|t|$. In the considered case of $i$-$wave$ altermagnetic state there are 12 regions with significant Fermi wave vector mismatch [cf. Fig. \ref{fig:triangular_cooper}(d)]. As shown in Fig. \ref{fig:triangular_pair_density} the non-zero momentum pairing allows to reduce the Fermi wave vector mismatch in 4 out of the 12 regions allowing to open the SC gap across a relatively large extent of the Fermi surface. It should be noted that the pure $d+id$ paired state appearing for low values of the $t_{\mathrm{am}}$ parameter is characterized by non-zero Chern number meaning that it has non-trivial topology. Nevertheless, for the FF phase obtained here, the topological features are lost.

For the case of a triangular lattice with the $i$-$wave$ altemagnetism, the general mechanism for the creation of the FF state is similar as in the case of the square lattice considered previously. Similarly as before, a mixed singlet-triplet state has been obtained with non-zero momentum of the Cooper pairs. However, it is difficult to interpret why the particular composition of the SC gap symmetry factors shown in Fig. \ref{fig:q_min_triangle}(d) has been obtained here as well as why at a certain critical value of the $t_{\mathrm{am}}$ parameter a transition appears which changes this composition. The difficulties in interpretation result from a more complex structure with a six-fold symmetry (instead of four-fold) meaning that during the optimization more degrees of freedom are available and therefore more possible symmetries of the SC gap parameter are allowed, as shown by Table \ref{tab:symmetries_SC}. Nevertheless, the general rule which is operative also here corresponds to reduction of the regions with significant Fermi wave vector mismatch and maximizing the SC gap at the Fermi surface. The subtle interplay between the mentioned factors most probably determines the effect of the transition obtained in Fig. \ref{fig:q_min_triangle}(c).

\section*{Conclusions}
We have carried out a comprehensive analysis of the interplay between unconventional superconductivity and altermagnetism in two-dimensional systems for the relevant cases of a square-lattice with $d$-$wave$ altermagnetic ordering  and a triangular-lattice with $i$-$wave$ altermagnetic ordering. We have taken into account all possible pairing symmetries and their mixtures that are allowed for the considered real-space pairing scenario. 

Our study shows that the stability of the FF phase with non-zero center-of-mass momentum of the Cooper pairs can appear in both considered altermagnetic states as a consequence of the Fermi wave-vactor mismatch resulting from the alternating spin splitting.
In particular, for the case of $d$-$wave$ altermagnetism, different pairing symmetries lead to different behaviors of the FF state. Namely, for the $extended$ $s$-$wave$ pairing, the FF phase with the $\mathbf{Q}$ vector oriented along the diagonal directions has been obtained, whereas for the $d$-$wave$ pairing the $\mathbf{Q}$ vector oriented in the horizontal and vertical directions lead to stability. This is a straightforward effect of the fully gapped (isotropic) situation realized in the $extended$ $s$-$wave$ paired state, for which the orientation of the Cooper pair momentum is determined purely by the symmetry of the alternating spin splitting. In such situation the diagonal $\mathbf{Q}$ vectors reduce the number of areas with significant Fermi wave vector mismatch from four to two allowing the gap to open across a significant extent of the Fermi surface. On the other hand, for the case of $d$-$wave$ paired state the appearance of the nodal directions in the SC gap changes the optimal orientation of the Cooper pair momentum. In such a case, the horizontal and vertical directions of the $\mathbf{Q}$ vector allow to avoid the nodal directions and still compensate for the Fermi wave vector mismatch at the same time. Additionally, a spin-triplet $p$-$wave$ admixture appears to the pairing symmetry, which moves slightly the nodal lines and optimize even more the SC gap distribution across the Fermi surfaces. Our calculations carried out in the presence of external magnetic field have shown that by changing the orientation of the magnetic field, one should be able to easily change the direction of the Cooper pair momentum in the FF state. Such an effect can be potentially interesting when it comes to various applications.  

As shown in the second part of our paper, a singlet-triplet mixing similar to that for the square lattice also appears for the case of triangular lattice with $i$-$wave$ altermagnetic symmetry. However, this time the obtained FF phase has a pattern of minima in the Cooper pair momentum space that conserves the $C_6$ symmetry. Furthermore, the SC gap parameter corresponds to a mixture of $d\pm id$ singlet and $f$-$wave$ triplet pairing. Additionally, by increasing the amplitude of the altrmagnetic spin splitting, we have reported a transition at which the composition of the particular SC gap symmetry components is changes. The $i$-$wave$ altermagnetic state leads to more degrees of freedom, allowing for significantly more complexity and exotic coexistent states in the context of interplay with superconductivity. 

The general rule which determines the features of the obtained FF states in all considered cases corresponds to adjusting the Cooper pair momentum as well as composing a suitable mix of pairing symmetry factors to open a maximally large SC gap across a significant extent of the Fermi surface. This allows for the stability of the paired phase in a seemingly unfavorable environment created by the presence of altermagnetic ordering. Subtle interplay between the mentioned factors leads to various types of anisotropic behaviors of the resultant non-zero momentum pairing, which has not been possible in the originally proposed FFLO\cite{Ferrell_FF,Larkin_LO} state. Interestingly, in the obtained paired phases additional symmetries appear, leading to exotic forms of the superconducting order parameter with singlet-triplet mixing.

Note that the smaller the SC gap, the more susceptible the paired state will be to the destructive influence of altermagnetic spin splitting. This leads to a situation in which the threshold value of t$_{\mathrm{am}}$ below which the SC state is stable approaches asymptotically zero with decreasing $J$. Nevertheless, even for relatively low values of $J$, non-zero threshold $t_{\mathrm{am}}$ can be obtained. However, this requires very precise and time-consuming calculations with a high number of $\mathbf{k}$-points taken for the integrals appearing in the self-consistent equations. Therefore, for our numerical analysis presented here, we have taken relatively large values of $J$ in order to perform the calculations in a reasonable time. Also, the effect of non-zero momentum pairing, which allows superconductivity to survive even above the threshold t$_{\mathrm{am}}$ is more robust for larger SC gaps. Therefore, promising candidates for the observation of the FFLO-type of states would be systems with high critical temperature, like the quasi-two-dimensional copper-based superconductors. In this context, one of the prominent members of the cuprate family is La$_2$CuO$_4$ which has been recognized to belong to a $d$-$wave$ altermagnetic spin group\cite{Smejkal_PRX_2,Smajkal_PRX_1}.

\bibliography{sample}

\section*{Acknowledgement}
This work is partly supported by the program „Excellence initiative – research university” for the AGH University of Krakow. We gratefully acknowledge the Polish high-performance computing infrastructure PLGrid (HPC Center: ACK Cyfronet AGH) for providing computer facilities and support within computational grant no. PLG/2024/017887 

\section*{Author contributions statement}
KJ has performed all the calculations presented in this work. MZ conceived the main idea behind this work and prepared the numerical code. KJ and MZ have wrote the first version of the manuscript. KJ, MZ, PW, and MN took part in the analysis and interpretation of the obtained results. All authors reviewed the manuscript.

\section*{Competing interests}
The authors declare no competing interests.

\section*{Data availability}
The datasets are available upon request from the corresponding author.

\section*{Code availability}
The codes are available upon request from the corresponding author.

%%%%%%%%%%%%%%%%%%%%%%%%%%%%%%%%%%%%%%%%%%%%%%%%%%%%%%%%%%%%%%%%%%%%%%%%%%%%%%%%%
\section*{Appendix A: Derivation of the Hamiltonian in $\mathbf{k}$-space}
Here we show in more detail how to derive the $\mathbf{k}$-space representation of the $\hat{H}_{sc}$ term. First, we rewrite the real space representation defined by Eq. (\ref{eq:H_SC_HF_real_space}) with explicit form of the gap amplitudes defined by Eq. (\ref{eq:Delta_ji}), namely, 
\begin{equation}
	\hat{H}_{sc} = -J \frac{1}{2}  \sum_{ij\sigma} \big[ \langle \hat{a}_{j\bar{\sigma}} \hat{a}_{i\sigma} \rangle \hat{a}_{i\sigma}^{\dagger}\hat{a}_{j\bar{\sigma}}^{\dagger}  + h.c \big] + \frac{1}{2J} \sum_{ij \sigma} |\Delta_{ji}^{\bar{\sigma}\sigma}|^2.
    \label{eq:H_SC_HF_real_space_Appendix}
\end{equation}
The trivial non-operator part we keep in the original form as we will not apply the transformation to it. Note that only nearest-neighboring terms are taken in the above summation since we neglect the longer-ranged pairings here. Next, we substitute the standard Fourier transform of the creation and anihilation operators, 
\begin{equation}
    \hat{a}_{i\sigma}=\frac{1}{\sqrt{N}}\sum_{\mathbf{k}}e^{-i\mathbf{k}\mathbf{R}_i}\;\hat{a}_{\mathbf{k}\sigma},\quad \hat{a}^{\dagger}_{i\sigma}=\frac{1}{\sqrt{N}}\sum_{\mathbf{k}}e^{i\mathbf{k}\mathbf{R}_i}\;\hat{a}^{\dagger}_{\mathbf{k}\sigma}.
\end{equation}
As a result we obtain
\begin{equation}\nonumber
\begin{split}
\hat{H}_{sc} &= -\frac{1}{2}J \sum_{ij\sigma}\frac{1}{N^2} \sum_{\mathbf{k}_1\mathbf{k}_2\mathbf{k}_3\mathbf{k}_4} \big[e^{i(\mathbf{k}_3-\mathbf{k}_2)\mathbf{R}_i} e^{-i(\mathbf{k}_1-\mathbf{k}_4)\mathbf{R}_j} \langle \hat{a}_{\mathbf{k}_1\bar{\sigma}} \hat{a}_{\mathbf{k}_2 \sigma} \rangle \hat{a}_{\mathbf{k}_3\sigma}^{\dagger} \hat{a}_{\mathbf{k}_4\bar{\sigma}}^{\dagger}+h.c. \big]  + \frac{1}{2J} \sum_{ij \sigma} |\Delta_{ji}^{\bar{\sigma}\sigma}|^2\\
&= -\frac{1}{2}J \sum_{ij\sigma}\frac{1}{N^2} \sum_{\mathbf{k}_1\mathbf{k}_2\mathbf{k}_3\mathbf{k}_4}\big[ e^{i(\mathbf{k}_3+\mathbf{k}_4-\mathbf{k}_1-\mathbf{k}_2)\mathbf{R}_i} e^{-i(\mathbf{k}_1-\mathbf{k}_4)(\mathbf{R}_j-\mathbf{R}_i)} \langle \hat{a}_{\mathbf{k}_1\bar{\sigma}} \hat{a}_{\mathbf{k}_2 \sigma} \rangle \hat{a}_{\mathbf{k}_3\sigma}^{\dagger} \hat{a}_{\mathbf{k}_4\bar{\sigma}}^{\dagger} + h.c.\big] + \frac{1}{2J} \sum_{ij \sigma} |\Delta_{ji}^{\bar{\sigma}\sigma}|^2,\\
\end{split}
\end{equation}
By introducing the Cooper pair momenta $\mathbf{Q}=\mathbf{k}_3 + \mathbf{k}_4$ and $\mathbf{Q}'=\mathbf{k}_1 + \mathbf{k}_2$ we obtain
\begin{equation}
\begin{split}
\hat{H}_{sc}&=-\frac{1}{2} J \sum_{\mathbf{k}_2\mathbf{k}_3\mathbf{Q}\mathbf{Q}'} \frac{1}{N^2} \sum_{ij}\Big[e^{-i(\mathbf{k}_3-\mathbf{k}_2)(\mathbf{R}_j-\mathbf{R}_i)}e^{i(\mathbf{Q}-\mathbf{Q}')\mathbf{R}_j} \langle \hat{a}_{(-\mathbf{k}_2+\mathbf{Q}')\bar{\sigma}} \hat{a}_{\mathbf{k}_2 \sigma}\rangle \hat{a}_{\mathbf{k}_3\sigma}^{\dagger} \hat{a}_{(-\mathbf{k}_3+\mathbf{Q})\bar{\sigma}}^{\dagger}+h.c.\Big]+ \frac{1}{2J} \sum_{ij \sigma} |\Delta_{ji}^{\bar{\sigma}\sigma}|^2\\
&=-\frac{1}{2} J \sum_{\mathbf{k}_2\mathbf{k}_3\mathbf{Q}\mathbf{Q}'}\Big[ \frac{1}{N^2} \Big(\sum_{j}e^{i(\mathbf{Q}-\mathbf{Q}')\mathbf{R}_j}\sum_{i}e^{-i(\mathbf{k}_3-\mathbf{k}_2)(\mathbf{R}_j-\mathbf{R}_i)}\Big) \langle \hat{a}_{(-\mathbf{k}_2+\mathbf{Q}')\bar{\sigma}} \hat{a}_{\mathbf{k}_2 \sigma}\rangle \hat{a}_{\mathbf{k}_3\sigma}^{\dagger} \hat{a}_{(-\mathbf{k}_3+\mathbf{Q})\bar{\sigma}}^{\dagger}+h.c.\Big]+ \frac{1}{2J} \sum_{ij \sigma} |\Delta_{ji}^{\bar{\sigma}\sigma}|^2,
\end{split}
\end{equation}
Note that in the above expressing the sum over $i$ in fact does not depends on $j$ due to translational symmetry of the triangular lattice. Therefore, the summation over $j$ and $i$ can be reorganised in the following manner
\begin{equation}
\begin{split}
\hat{H}_{sc}&=-\frac{1}{2} J \sum_{\mathbf{k}_2\mathbf{k}_3\mathbf{Q}\mathbf{Q}'}\Big[ \frac{1}{N^2} \Big(\sum_{j}e^{i(\mathbf{Q}-\mathbf{Q}')\mathbf{R}_j}\Big)\Big(\sum_{i(j)}e^{-i(\mathbf{k}_3-\mathbf{k}_2)(\mathbf{R}_j-\mathbf{R}_i)}\Big) \langle \hat{a}_{(-\mathbf{k}_2+\mathbf{Q}')\bar{\sigma}} \hat{a}_{\mathbf{k}_2 \sigma}\rangle \hat{a}_{\mathbf{k}_3\sigma}^{\dagger} \hat{a}_{(-\mathbf{k}_3+\mathbf{Q})\bar{\sigma}}^{\dagger}+h.c.\Big]+ \frac{1}{2J} \sum_{ij \sigma} |\Delta_{ji}^{\bar{\sigma}\sigma}|^2,
\end{split}
\end{equation}
where the notation $i(j)$ means that the summation runs over the nearest-neighboring lattice sites $i$ surrounding the site $j$ (pairing appears only between neighboring lattice sites). After using the Dirac delta represented in the form $\delta(\mathbf{Q}-\mathbf{Q}')=\sum_je^{i(\mathbf{Q}-\mathbf{Q'})\mathbf{R}_j}/N$, we can rewrite the above expression as follows
\begin{equation}
    \hat{H}_{sc}=\frac{1}{2}\sum_{\mathbf{k}\mathbf{Q}\sigma}\Big[\Delta^{\bar{\sigma}\sigma}_{\mathbf{k}\mathbf{Q}}\;\hat{a}_{\mathbf{k}\sigma}^{\dagger} \hat{a}_{(-\mathbf{k}+\mathbf{Q})\bar{\sigma}}^{\dagger}+h.c.\Big]+ \frac{1}{2J} \sum_{ij \sigma} |\Delta_{ji}^{\bar{\sigma}\sigma}|^2,
    \label{eq:derivation_finish}
\end{equation}
where 
\begin{equation}
    \Delta^{\bar{\sigma}\sigma}_{\mathbf{k}\mathbf{Q}} = \sum_{i(j)}e^{i\mathbf{k}(\mathbf{R}_i-\mathbf{R}_j)} \Delta_{ji\mathbf{Q}}^{\bar{\sigma}\sigma},
\end{equation}
and
\begin{equation}
    \Delta_{ji\mathbf{Q}}^{\bar{\sigma}\sigma} = -\frac{J}{N} \sum_{\mathbf{k}}e^{-i\mathbf{k}(\mathbf{R}_i-\mathbf{R}_j)} \langle \hat{a}_{(-\mathbf{k}+\mathbf{Q})\bar{\sigma}} \hat{a}_{\mathbf{k} \sigma}\rangle.
    \label{eq:Delta_ij_app}
\end{equation}
Additionally, the non-operator part can be expressed with the use of $\Delta_{ij\mathbf{Q}}^{\bar{\sigma}\sigma}$ in the following manner
\begin{equation}
    \frac{1}{2J} \sum_{ij \sigma} |\Delta_{ji}^{\bar{\sigma}\sigma}|^2=\frac{N}{2J} \sum_{\mathbf{Q}}\sum_{i(j) \sigma} |\Delta_{ji\mathbf{Q}}^{\bar{\sigma}\sigma}|^2
\end{equation}
As one can see in Eq. (\ref{eq:derivation_finish}) pairing between $(\mathbf{k},\;\sigma)$ and $(-\mathbf{k}+\mathbf{Q},\; \bar{\sigma})$ electrons is taken into account for all possible momenta of the Cooper pairs $\mathbf{Q}$. For the most trivial case with no spin-splitting, the ground-state of the Cooper pair condensate corresponds to $\mathbf{Q}=0$. When it comes to non-zero momentum pairing induced by spin splitting, there are two basic variants. The first corresponds to the Fulde-Ferrell state (FF)\cite{Ferrell_FF} with only one vector of the Cooper pair momentum ($\mathbf{Q}$), and the second is the Larkin-Ovhinnikov state (LO)\cite{Larkin_LO} with two opposite vectors of the Cooper pair momenta ($\mathbf{Q}$ and $-\mathbf{Q}$). For the sake of simplicity, in the main text we consider the FF state, for which the summation over the $\mathbf{Q}$ vector can be dropped since a single $\mathbf{Q}$ is considered. Within our approach, the Cooper pair momentum $\mathbf{Q}$ is determined by minimizing the energy of the system.

\end{document}